\documentclass[12pt]{article}
\usepackage{amssymb}
\usepackage{amsfonts}
\usepackage{youngtab}
\usepackage{amsmath,amssymb,latexsym,cite}
\usepackage{amsthm}
\usepackage{cite}
\usepackage[a-1b]{pdfx}    % archival
\usepackage[]{hyperref}
\input xy
\xyoption{all}

\numberwithin{equation}{section}

\def\P{{\mathbb{P}}}
\def\R{{\mathbb{R}}}
\def\Z{{\mathbb{Z}}}
\def\C{{\mathbb{C}}}
\def\cO{{{\cal O}}}
\def\cR{{{\cal R}}}
\def\cH{{{\cal H}}}

\def\ra{{{\rangle}}}

\def\Tr{{\text{Tr}}}
\def\tb{{\overline{t}}}

\begin{document}

%\vspace*{0.5in}

\begin{center}

{\large\bf Quantum cohomology from mixed Higgs-Coulomb phases}

\vspace*{0.2in}

Wei Gu$^1$, Ilarion~V.~Melnikov$^2$, Eric Sharpe$^3$ 

\begin{tabular}{cc}
{\begin{tabular}{l}
$^1$ Max-Planck-Institut f\"ur Mathematik\\
Vivatsgasse 7\\
D-53111 Bonn, Germany
\end{tabular}}
&
{\begin{tabular}{l}
$^2$ Department of Physics and Astronomy\\
James Madison University \\
Harrisonburg, VA 22807
\end{tabular}}
\end{tabular}

\begin{tabular}{c}
{\begin{tabular}{l}
$^3$ Department of Physics MC 0435\\
850 West Campus Drive\\
Virginia Tech\\
Blacksburg, VA  24061 \end{tabular}}
\end{tabular}

{\tt weig8@vt.edu}, 
{\tt melnikix@jmu.edu},
{\tt ersharpe@vt.edu}

$\,$

\end{center}

We generalize Coulomb-branch-based gauged linear sigma model
(GLSM)--computations of
quantum cohomology rings of Fano spaces.
Typically such computations have focused on GLSMs without superpotential,
for which the IR phase of the GLSM is a pure
Coulomb branch, and quantum cohomology is determined by the critical
locus of a twisted one-loop effective superpotential.
Here, we systematically extend to cases for which
the IR phase is a mixture of Coulomb and Higgs branches, where the latter is a
Landau-Ginzburg orbifold.
We describe the state spaces and products of corresponding operators in detail,
comparing a geometric phase description, where the OPE ring is quantum
cohomology,
to the IR description
in terms of Coulomb and Higgs branch states.
As a concrete test of our methods,
we compare to existing mathematics results for quantum cohomology rings
of hypersurfaces in projective spaces in numerous examples.

\begin{flushleft}
August 2023
\end{flushleft}

\newpage

\tableofcontents

\newpage

\section{Introduction}

Quantum cohomology is a deformation of an
ordinary cohomology of a space $X$, which is computed mathematically 
via an analysis
of intersection theory on the moduli space of worldsheet instantons on $X$,
and interpreted physically in terms of OPEs of operators in A-twisted
nonlinear sigma models with target $X$.

If the nonlinear sigma model for $X$ can be described as a phase of a gauged linear sigma model (GLSM)\cite{Witten:1993yc}, then its quantum cohomology can also be determined in other phases, which may 
correspond to different geometries or have a non-geometric description.
This is possible because the different phases are in fact smoothly connected in the deformation space of
the GLSM,\footnote{
This is true even in non-Calabi-Yau cases.  Although the axial R symmetry
is anomalous, and hence the theta angle not well-defined for all values,
there is a discrete nonanomalous subgroup, as we shall utilize later,
and so some subset of the theta angle survives, which is sufficient for
the analytic continuation arguments of \cite{Witten:1993yc}.
} and it should be possible to perform computations
of topological observables in the different descriptions.
For example, the GLSM A-model correlation functions are meromorphic functions of the complexified K\"ahler parameters, and it should be possible to calculate them in any particular phase.  This was confirmed for Calabi-Yau hypersurface examples in~\cite{Morrison:1994fr} by summing the gauge instanton sums.  In that case a choice of phase determines the  topological sectors that contribute to the correlation functions, and while each sum has a different region of convergence in the space of complexified K\"ahler parameters, they all lead to the same set of meromorphic correlation functions.

When $X$ is a Fano space, then the GLSM's complexified Fayet-Iliopoulos  (FI) parameters have
a non-trivial renormalization group (RG) running.  In the geometric phase, where the classical low energy
physics of the GLSM is that of a nonlinear sigma model with target $X$, these parameters map to the
complexified K\"ahler parameters of the nonlinear sigma model.
Since the topological sector of the theory is RG--invariant, 
we expect quantum cohomology relations in the different phases
to match.
For example, if Fano $X$ can be realized as a phase of
GLSM without superpotential,
meaning for example a Fano toric variety, Grassmannian, or flag
manifold, then under renormalization group flow, the geometric
phase flows to a pure Coulomb branch phase, where the quantum cohomology
relations can be computed
using purely algebraic methods as the critical locus of a one-loop
twisted effective superpotential \cite{Morrison:1994fr}.
For Fano spaces $X$ described by GLSMs without a superpotential,
these methods are by now standard in the community.

The goal of this paper is to extend the previous analysis to situations
in which the 
UV phase of the GLSM describes a Fano hypersurface in a toric variety.  
In such theories,
the IR phase typically includes both Coulomb and Higgs vacua.  
We describe the computation of quantum cohomology rings in this
more complicated setting, in mixed Higgs-Coulomb branches.
Extensive mathematics results on quantum cohomology rings exist for such
cases, and we check our computations by comparing to those results.

The theories we study have features common to both the case of toric and Calabi-Yau GLSMs:  like the former they feature an IR phase with Coulomb vacua, and like the latter they have a non-trivial IR Higgs sector.  On the other hand, they also have some slightly unusual features.  For example, the Higgs sector of the IR theory has an emergent superconformal symmetry, and the matching of the UV and IR descriptions of the A-model require us to combine contributions from the Coulomb and Higgs sectors in the IR.
We shall focus on the prototypical example of hypersurfaces in projective
spaces, which will already exhibit a rich structure and nontrivial matchings.

We begin in section~\ref{sect:basic} by outlining the basic ideas behind
the mixed Higgs-Coulomb branch computations we describe in this paper.
We will extensively test our methods in GLSMs describing hypersurfaces
in projective spaces, as the mathematics community has extensive
results for both quantum cohomology rings and Gromov-Witten
invariants 
in 
complete intersections in projective spaces
(see e.g.~\cite{beauville,sheridan,abpz,Collino:1996my}).
To that end, in section~\ref{sect:rev-math} we review pertinent mathematics
results, against which we will compare later.
In section~\ref{sect:phys} we describe our physical computations.
The vector space structure underlying the quantum cohomology rings in the 
IR phase will arise from
a combination of both Coulomb branch states as well as Landau-Ginzburg
orbifold states, and we describe the computation of spectra and
give general arguments explaining why those spectrum computations and
the operator products should reproduce the known mathematics results
reviewed earlier.  To further convince the reader,
in section~\ref{sect:exs} we check our methods and computations
by applying them to numerous concrete
examples of hypersurfaces in projective spaces.
We emphasize that our methods apply in generality.

Finally, in section~\ref{sect:other} we outline attempts to apply these
same methods to two other analogous quantities which can be computed
in GLSMs, namely quantum K theory and quantum sheaf cohomology.
In both cases, the methods we have described so far do not completely suffice
to fully capture the rings, but we do capture some of the structure.
We leave the complete determination of the rings in those cases for the future.

Let us take a moment to set our conventions.
In this paper, we still study (A-twisted)
gauged linear sigma models (GLSMs) describing
Fano manifolds, usually hypersurfaces ${\mathbb P}^n[d]$ for $d \leq n$.
In a GLSM for a non-Calabi-Yau space, the Fayet-Iliopoulos parameter\footnote{
We note that the complexified Fayet-Iliopoulos
parameters represent the complexified K\"ahler parameters in algebraic coordinates.
While this is the natural choice of coordinates from the GLSM perspective, in general
these are related in a non-trivial fashion to the natural coordinates of the nonlinear model:  see 
the discussion around equation (4.20) in~\cite{Morrison:1994fr}, for example. 
}
$r$ has nontrivial RG flow, and will flow between different phases of the
GLSM.  In our examples there are two phases: the
``UV phase,'' or ``geometric phase,'' for which $r \gg 0$,
which corresponds to the geometry ${\mathbb P}^n[d]$,
and the ``IR phase,'' for which $r \ll 0$,
which consists of a combination of Coulomb vacua
and a Higgs branch given by a Landau-Ginzburg orbifold.   This
generalizes the well-known Calabi-Yau/Landau-Ginzburg correspondence,
in which case there are two related simplifications:  there is no RG flow,
and the $r\ll 0$ phase has no Coulomb vacua.  For Fano hypersurfaces, 
Coulomb vacua are present,
and the IR phase has both Higgs and Coulomb sectors, which decouple
in the topological theory.

The Coulomb vacua will be described
in terms of the scalar component $\sigma$ of the twisted chiral superfield
$\Sigma$ defined by \cite[equ'n (2.16)]{Witten:1993yc}
\begin{equation}
\Sigma \: = \: 
\frac{1}{\sqrt{2}} \overline{D}_+ D_- V \: = \: \sigma \: + \: \cdots,
\end{equation}
where $V$ is a vector superfield, as we describe later.
For a hypersurface ${\mathbb P}^n[d]$, the Higgs branch of the
IR phase is described by a ${\mathbb Z}_d$ orbifold of a Landau-Ginzburg
model on ${\mathbb C}^{n+1}$ with a superpotential $W$ that is homogeneous
of degree $d$, given by the defining hypersurface.

\section{A central decoupling observation}
\label{sect:basic}

For Fano toric varieties, the quantum cohomology ring relations were computed in
GLSMs in \cite{Morrison:1994fr} in two ways:
by summing the gauge instantons in the UV geometric phase,
 and also from the critical locus of
the one-loop twisted 
effective superpotential for the $\Sigma$ multiplets in the IR phase.  
For hypersurfaces and
complete intersections, the IR phase is no longer a pure Coulomb branch,
but is instead a mixture of Coulomb and Higgs branches, and both must be
taken into account to reproduce the quantum cohomology structure of the
UV phase.

One central idea behind our computations is that deep in the
IR, the Coulomb and Higgs branches are disconnected.
Correlation functions are computed\footnote{
On connected worldsheets, which we assume for simplicity.
} as a sum of Coulomb branch computations
(with vanishing Higgs fields) and Higgs branch computations (with
vanishing $\sigma$ fields).  This is ultimately a consequence of certain
GLSM bosonic potential
terms which in a $U(1)$ GLSM have the form
\begin{equation}
| \sigma |^2 \sum_i Q_i^2 | \phi_i |^2,
\end{equation}
for $\phi_i$ the scalar components of the matter chiral superfields.
As a result,
\begin{itemize}
\item If the $\phi_i$ are nonzero and large, then $\sigma$ is massive,
hence we take $\sigma = 0$, corresponding to the Higgs branch,
\item If $\sigma$ is large, then the Higgs fields are very massive,
corresponding to the Coulomb branch,
\end{itemize}
and in any event, both $\sigma$ and $\phi_i$ are
not simultaneously nonzero at low energies.
As a result, 
in mixed Higgs-Coulomb phases,
if $t$ represents a Landau-Ginzburg orbifold field and $\sigma$ a Coulomb
branch field, then
\begin{equation}  \label{eq:root}
\sigma \cdot t \: = \: 0.
\end{equation}
This is an implementation of the `reliability criterion' described
in \cite{Melnikov:2005hq,Melnikov:2006kb}, and we will see that this
appears as one of the relations defining OPEs and quantum cohomology
rings.

We will combine equation~(\ref{eq:root}) 
and Coulomb branch relations for the $\sigma$
field, derived as in \cite{Morrison:1994fr} from a one-loop-exact
twisted effective superpotential, 
with an analysis of the Landau-Ginzburg orbifold states to
give an IR phase description of 
the quantum cohomology ring relations for Fano hypersurfaces in
projective space.  We will see that although Higgs and Coulomb branches
are decoupled, the Dolbeault cohomology is represented by a linear combination
of contributions from both.

\section{Review of pertinent mathematics}
\label{sect:rev-math}

The quantum cohomology rings of Fano complete intersections in projective spaces
have been
well-studied mathematically,
see e.g.~\cite{beauville}, \cite[prop. 1.13]{sheridan},
\cite[theorem D]{abpz}, \cite{Collino:1996my}. 
The point of this paper
is to understand the corresponding physics computations.
In this section we review known mathematical results for these spaces,
against which we will compare physics results later.

Thanks to the Lefschetz hyperplane theorem,
the cohomology of a complete intersection
${\mathbb P}^n[d_1, \cdots, d_m]$ of all degrees below $n-m$ and above
$n+m$
is inherited from that of ${\mathbb P}^n$.
In general, there can be additional states
in middle degrees, whose index can be computed from the Euler characteristic
of the entire space.

For example, for a hypersurface ${\mathbb P}^n[d]$, there are possibly
extra states in degree $n-1$, which can be determined
from\footnote{
This can be computed by, for example, computing the top Chern class of the
tangent bundle from the sequence defining the tangent sheaf of $X$, and evaluating
on the fundamental class of the hyperplane.
}
\begin{equation}  \label{eq:eulerchar}
\chi( {\mathbb P}^n[d] ) \: = \:
\sum_{i=0}^{n-1} \left( \begin{array}{c} n+1 \\ i \end{array} \right)
(-1)^{n-1-i} d^{n-i}~.
\end{equation}
As a result, we see that the primitive cohomology has dimension
\begin{equation}  \label{eq:total-lgstates}
|\chi({\mathbb P}^n[d])| - n \: = \: \left|
\sum_{i=0}^{n-1} \left( \begin{array}{c} n+1 \\ i \end{array} \right)
(-1)^{n-1-i} d^{n-i}\right| \: - \: n~.
\end{equation}

In the quantum field theory quantum cohomology arises as the algebra of local operators of an A-twisted non-linear sigma model associated to the Fano variety $X$.  The space of local operators in such a theory is isomorphic to the de Rham cohomology $H^{\bullet}(X,\R)$, and to each class $[\omega] \in H^k(X,\R)$ we associate an operator in the A-model denoted by ${\cal O}[\omega]$.

Now, the quantum cohomology and Gromov-Witten theory of Fano complete
intersections in projective spaces has been described mathematically in
e.g.~\cite{beauville},
\cite[prop. 1.13]{sheridan},
\cite[theorem D]{abpz}, \cite{Collino:1996my}.
As stated in \cite{beauville},
the quantum cohomology ring of a Fano 
complete intersection ${\mathbb P}^n[d_1, \cdots, d_m]$ is
generated by
the hyperplane class $\eta$ and the primitive cohomology elements
$\alpha \in H^{n-m}_{\text{prim}}(X,{\mathbb Q})$ with relations\footnote{
The reader should note that reference~\cite{beauville} discusses
complete intersections in ${\mathbb P}^{n+m}$, whereas we work
in ${\mathbb P}^n$, so our ``$n$'' differs from that of \cite{beauville}.  
We have also normalized $q$ slightly differently,
to be consistent with conventions elsewhere in this paper based on the
natural GLSM coordinate.
}
\begin{equation}  \label{eq:math:qcci1}
\eta^{n+1-m} = q \left( \prod_{a=1}^m \left(-d_a \right)^{d_a} \right)
\eta^{\sum_a (d_a - 1)}, \: \: \: 
\eta \alpha = 0, 
\end{equation}
\begin{equation}  \label{eq:math:qcci2}
\alpha \cdot \beta = (\alpha | \beta) \frac{1}{d}
\left( \eta^{n-m} - q \left( \prod_{a=1}^m  \left( - d_a \right)^{d_a}  \right)
\eta^{\sum_a (d_a - 1) - 1}  \right),
\end{equation}
for\footnote{
This restriction was included in the statement of the theorem in
\cite{beauville}, but our results apply to any Fano complete intersection,
and indeed we will study several examples that lie outside this bound.
}
\begin{equation}  \label{eq:n-constr}
n - m \: \geq \: 2 \sum_a \left( d_a - 1 \right) - 1,
\end{equation}
where
$\alpha, \beta \in H^{n-m}_{\text{prim}}(X,{\mathbb Q})$,
$(\alpha | \beta)$ is the ordinary classical product in
middle cohomology, the ``prim'' subscript indicates that $\alpha, \beta$
lie in primitive cohomology, meaning they are annihilated in cohomology
by the hyperplane class $\eta$:
\begin{equation}
[\alpha \wedge \eta] \: = \: 0 \: = \: [\beta \wedge \eta],
\end{equation}
and, finally, $d$ is the degree of the complete intersection:
\begin{equation}
d \: = \: \prod_{a=1}^m d_a.
\end{equation}
We shall see that it is no accident that the second relation
in~(\ref{eq:math:qcci1}) resembles the decoupling relation~(\ref{eq:root})
described earlier.

As a consistency check, the reader should note that the
ring relations above are linked.  For example, 
from cohomological degrees, it must be the case that
\begin{equation}
\alpha \cdot \beta \: \propto \:
\eta^{n-m}, \:
q \eta^{ \sum_a (d_a - 1) - 1}.
\end{equation}
Furthermore, from~(\ref{eq:math:qcci1}) (or, ultimately from the
physics relation~(\ref{eq:root})), 
\begin{equation}
\eta \cdot \alpha \cdot \beta \: = \: 0,
\end{equation}
hence the linear combination must be in the kernel of $\eta$, so
from the first part of~(\ref{eq:math:qcci1}), we see
\begin{equation}
\alpha \cdot \beta \: \propto \:
 \eta^{n-m} - q \left( \prod_{a=1}^m  \left( - d_a \right)^{d_a}  \right)
\eta^{\sum_a (d_a - 1) - 1}.
\end{equation}

In terms of quantum field theory computations, 
the quantum cohomology ring is generated by $\cO[\eta]$, 
the operator corresponding to the hyperplane class $\eta$ as well as $\cO[\alpha]$, where $\alpha$ belong to the primitive cohomology
$H^{n-m}_{\text{prim}}(X,{\mathbb Q})$, which obey relations corresponding
to the above:
\begin{equation}  \label{eq:phys:qcci1}
(\cO[\eta])^{n+1-m} = q \left( \prod_{a=1}^m \left(-d_a \right)^{d_a} \right)
(\cO[\eta])^{\sum_a (d_a - 1)}~, \: \: \: 
\cO[\eta] \cdot \cO[\alpha] = 0~, 
\end{equation}
\begin{equation}  \label{eq:phys:qcci2}
\cO[\alpha] \cdot \cO[\beta] = (\alpha | \beta) \frac{1}{d}
\left( (\cO[\eta])^{n-m} - q \left( \prod_{a=1}^m  \left( - d_a \right)^{d_a}  \right)
(\cO[\eta])^{\sum_a (d_a - 1) - 1}  \right).
\end{equation}
In the UV phase this is just a restatement of the mathematical results, emphasizing the difference between classical cohomology classes and the corresponding topological field theory operators.  Our main goal is to understand how these relations arise in the IR phase.  We will see that while the second equation in~(\ref{eq:phys:qcci1}) is a consequence of the decoupling between the Higgs and Coulomb branches sketched in~(\ref{eq:root}), the remaining equations involve a more intricate interplay between the Coulomb and Landau-Ginzburg orbifold fields.

We will recover these equations from detailed considerations in the
next section.
Specifically, equation~(\ref{eq:phys:qcci1}) 
will appear later as equations~(\ref{eq:final:c1}), (\ref{eq:final:root}),
and equation~(\ref{eq:phys:qcci2}) will appear later as
equations~(\ref{eq:final:lg1}).
We will also see a nontrivial mixing between Higgs and Coulomb contributions.
For example, although~(\ref{eq:phys:qcci1}) naively appears to solely
result from Coulomb branch computations, later we will see that
${\cal O}[\eta]$ is a linear combination of both Coulomb and Higgs
branch contributions.

The dimension of the primitive middle cohomology can be obtained from an 
index computation, as in equation~(\ref{eq:total-lgstates}). For our purposes it will be important to 
refine the structure further and describe the operators in terms of 
Dolbeault cohomology $H^{k,\ell}(X)$.  
In the case of a degree $d$ hypersurface in $\P^{k+\ell+1}$ 
the dimensions of the primitive Dolbeault cohomology groups can be obtained from an elegant generating function
\cite[chapter 17, theorem 17.3.4]{arapura}
\begin{equation}  \label{eq:gen-fn-hodge}
\sum_{k,\ell} \left( h^{k,\ell} - \delta_{k,\ell} \right)
x^k y^{\ell} \: = \:
\frac{
(1+y)^{d-1} - (1+x)^{d-1}
}{
(1+x)^d y - (1+y)^d x
}.
\end{equation}

If we let $D_p(d,n-1)$ denote the
dimension of the primitive degree-$(n-1)$ 
cohomology of a degree $d$ hypersurface
in ${\mathbb P}^n$, then from the generating function above,
\begin{equation}  \label{eq:midcoh:dim}
D_p(n,d) \: = \: 
\frac{ (d-1)^{n+1} + (-)^{n} }{d} \: + \: (-)^{n-1}.
\end{equation}

\section{Physical computations for hypersurfaces}
\label{sect:phys}

In this section we will describe the mixed Higgs-Coulomb branch computations
which give a physical realization of Dolbeault cohomology $H^{k,\ell}$.
Although our initial analysis applies to complete intersections in
projective spaces, we will quickly specialize to just hypersurfaces.
Briefly, for the IR phase corresponding to
a hypersurface of degree $d$ in ${\mathbb P}^n$,
\begin{itemize}
\item The Coulomb branch will contribute $n+1-d$ states which will correspond
to elements of $H^{k,k}$.
\item The Higgs branch is a ${\mathbb Z}_d$ Landau-Ginzburg orbifold (LGO),
with superpotential $W$ given by the homogeneous polynomial
defining the hypersurface.
The twisted sectors (indexed by $0 \leq r < d$) contribute states as follows:
\begin{itemize}
\item $r=0$: one state contributing to $H^{k,k}$ for $2k \equiv 0 \mod 2(n+1-d)$.
\item $r=1$: For each $p$ such that $p+n+1 \equiv 0 \mod d$, as many states
as the dimension of the space of degree $p$ polynomials modulo the ideal
$(dW)$, contributing to $H^{k,\ell}$ for
\begin{equation}
\ell-k \: = \: \frac{1}{d} \left( 2 (p+n+1) - d(n+1) \right),
\: \: \:
k+\ell \equiv n-1 \mod 2(n+1-d),  
\nonumber
\end{equation}
\item $1<r<d$: one state for each $r$, contributing to $H^{k,k}$  for
$2k \equiv 2(n+1-r) \mod 2(n+1-d)$.
\end{itemize}
\end{itemize}
The identification of states with elements of Dolbeault cohomology utilizes
a nonanomalous vector $U(1)_R$ symmetry and a nonanomalous
${\mathbb Z}_{2(n+1-d)}$ axial R-symmetry. 
Because one of these is a finite group, the identification of states with elements
of Dolbeault cohomology is ambiguous --
some elements of Dolbeault cohomology
can only be identified with linear combinations of Coulomb and Higgs branch
states.

We will argue that these states completely span the quantum cohomology
ring.  In subsequent sections, we will check the details in examples.

\subsection{General idea}

Consider a complete intersection of
hypersurfaces of degrees $d_1, d_2, \cdots, d_m$ in ${\mathbb P}^n$.
Physically, this corresponds to a $U(1)$ GLSM with $n+1$
fields $x_i$ of charge $+1$, $m$ fields $p_a$ of charge $-d_a$,
and a superpotential
\begin{equation}
W \: = \: \sum_{a=1}^m p_a G_a(x_i),
\end{equation}
where $G_a(x_i)$ is a homogeneous polynomial of degree $d_a$.

In the case
\begin{equation}
\sum_a d_a \: < \: n+1,
\end{equation}
the $r \gg 0$ phase (known henceforward as the UV phase or
geometric phase) describes
a Fano complete intersection, and the $r \ll 0$
phase of this theory (known henceforward as the IR phase)
is a mixed Higgs-Coulomb phase,
consisting of the IR limit of the (possibly hybrid) Landau-Ginzburg
orbifold (expected to be a nontrivial SCFT), and a decoupled set of
Coulomb vacua.  The theory flows to this phase in the IR.

In the UV (geometric) phase 
(meaning, the phase describing a large-radius sigma model on
${\mathbb P}^n[d]$), the state space corresponds (additively) to
Dolbeault cohomology.
We integrate out the gauge fields, and we are left with a non-linear sigma model with target space $X$.  
Its local A-model observables are then exactly described by operators 
$\cO[\eta]$ (corresponding to the hyperplane class)
and $\cO[\alpha]$ (corresponding to primitive cohomology in middle degree).  
On the other hand, in the IR phase (meaning
the mixed Coulomb / Landau-Ginzburg orbifold theory),
these should have a description in terms of the field $\sigma$ and fields associated to a Higgs (Landau-Ginzburg orbifold) phase where $\sigma =0$.  

For a general complete intersection the physics of the Higgs phase is that of a hybrid 
Landau-Ginzburg orbifold (LGO) theory---a rather complicated quantum field theory involving a LG sector coupled to a non-linear sigma model sector~\cite{Bertolini:2013xga,Bertolini:2014dia,Bertolini:2018now,Guo:2021aqj,Erkinger:2022sqs}.  Fortunately, for a hypersurface the Higgs phase is a much simpler LGO model which can be studied using the techniques developed  in~\cite{Vafa:1989xc,Intriligator:1990ua,Kachru:1993pg}.   This is the case to which we will specialize and show how to match the large radius geometric UV phase description to that of the IR phase.

Applying the methods of \cite{Morrison:1994fr} to the Coulomb branch,
from the one-loop-exact twisted effective superpotential
in this model, the $\sigma$ field obeys
\begin{equation}   \label{eq:qc:ci}
\sigma^{n+1-m} \: = \: q \left( \prod_{a=1}^m \left( - d_a \right)^{d_a}
\right) \sigma^{\sum_a (d_a - 1) }.
\end{equation}
Since in the UV (large radius) phase $\sigma$ can be interpreted 
as $\cO[\eta]$ (the operator associated to the hyperplane class), 
it may seem that the IR Coulomb branch already appears to match~(\ref{eq:math:qcci1}),
and given the proposed OPE~(\ref{eq:root}), we just have to understand~(\ref{eq:math:qcci2}) from a physical perspective.
However, as we will see, even this is far from a complete story because the Coulomb branch only contributes $n+1-d$ states, and when $d>1$ these do not span all of $H^{k,k}(X)$.  Instead, the Landau-Ginzburg orbifold sector provides the additional states necessary to match both the additional vertical cohomology and the primitive cohomology of $X$.

The product structure on the quantum cohomology ring is also more involved.  To describe it
we decompose the IR states (and therefore also the corresponding fields) into three
classes:
\begin{enumerate}
\item the Coulomb branch vacua;
\item the first twisted sector of the Landau-Ginzburg orbifold;
\item the remaining $d-1$ sectors of the Landau-Ginzburg orbifold.
\end{enumerate}
We will argue in section~\ref{sect:matching} that while the second class corresponds to the primitive cohomology on $X$,
the first and third correspond to the remaining vertical cohomology groups $H^{k,k}(X)$.  By studying the spectrum and symmetries of the A-model in the IR phase, we find a field in the IR description that corresponds to the UV field $\cO[\eta]$, 
as well as the IR fields $\Xi[\alpha]$ that map to the $\cO[\alpha]$ of the UV description, and these fields have OPE structure compatible with (\ref{eq:math:qcci1}) and~(\ref{eq:math:qcci2}).  Our proposal is summarized in equation~(\ref{eq:UVIRmap}).

\subsection{Hypersurface GLSM symmetries}

Specializing to hypersurfaces $\P^n[d]$, we would now like to match the A-model local observables or, by the state--operator correspondence in a topological field theory~\cite{Witten:1990bs,Dijkgraaf:1990qw}, the states between the UV and IR descriptions.  To facilitate the match it will be useful to consider the symmetries of the GLSM.  In this section we will identify the pertinent symmetries and describe their action on the local A-model observables.

Classically, the GLSM Lagrangian has a $U(1)_{L} \times U(1)_{R}$ symmetry which acts on the superspace coordinates $\theta^-$ and $\theta^+$ with charges $(1,0)$ and $(0,1)$, respectively, while assigning charge $(1,1)$ to the chiral superfield $\Phi$ and leaving the $X_i$ neutral.  We will denote the charges of a state with respect to these symmetries by $q_L$ and $q_R$.  

The non-chiral or ``vector'' $U(1)$ symmetry which assigns charge $q_V = q_L+q_R$ remains a symmetry of the quantum theory, while the chiral $U(1)$ symmetry is anomalous.  Fortunately, there is a non-anomalous $\Z_{2(n+1-d)}$ subgroup with generator $\rho$ that acts by
\begin{align}
\rho \cdot \cO_{q_L,q_R} = \zeta^{q_R-q_L} \cO_{q_L,q_R}~,
\end{align}
where
\begin{align}
\zeta = \exp\{ i\pi /(n+1-d)\}~.
\end{align}
In a little more detail, the action of $\rho$ on the superfields and their components is as follows:
\begin{equation}
\rho \cdot \theta^+ \: = \: \zeta^{-1} \theta^+, \: \: \:
\rho \cdot \theta^- \: = \: \zeta \theta^-, \: \: \:
\rho \cdot \Phi \: = \: \Phi, \: \: \:
\rho \cdot X_i \: = \: X~,
\end{equation}
or on components by charges as follows:
\begin{center}
\begin{tabular}{cc|cc|cc|cc}
Field & Charge & Field & Charge & Field & Charge & Field & Charge \\ \hline
$x$,~$\phi$ & $0$ & $\overline{x}$,~$\overline{\phi}$ & $0$ & $\sigma$ & $+2$ & $\overline{\sigma}$ & $-2$
\\
$\psi_+$ & $-1$ & $\overline{\psi}_+$ & $+1$ & $\lambda_+$ & $-1$ & 
$\overline{\lambda}_+$ & $+1$ \\
$\psi_-$ & $+1$ & $\overline{\psi}_-$ & $-1$ & $\lambda_-$ & $+1$ &
$\overline{\lambda}_-$ & $-1$
\end{tabular}
\end{center}

Using these symmetries we can characterize the operators in the UV phase
and the corresponding elements of Dolbeault cohomology.  
Given a class $[\omega] \in H^{k,\ell}(X)$,
the operator $\cO[\omega]$ carries charge 
\begin{equation} \label{eq:dol-1}
q_V \: = \: q_L + q_R \: = \: \ell - k,
\end{equation}
and $\rho\cdot \cO[\omega] = \zeta^{k+\ell} \cO[\omega]$.  We also see that the operator $\sigma$ carries the same charges as $\cO[\eta]$, as it should be.  For each of these local operators we also obtain a corresponding state in the A-model, which we denote as $|\omega;\text{uv}\rangle$, so that the A-model Hilbert space has a UV presentation
\begin{align}
\cH_{\text{uv}} \simeq \text{Span}\{ \cO[\eta]^k |\Omega;\text{uv}\ra,~~k=0,\ldots,n-1,~~  \cO[\alpha]|\Omega;\text{uv}\ra~,\alpha \in H^{n-1}_{\text{prim}}(M)\}~.
\end{align}
We will give a more explicit description of the $\rho$ action on
Landau-Ginzburg orbifold states in 
equation~(\ref{eq:rho-action}).

\subsection{Hyperplane class correlation functions}

In this section we will perform localization computations of correlation functions
of operators corresponding to powers of the hyperplane class, in the
UV and IR, using the GLSM fields.  We will see that they match each other in a non-trivial fashion and reproduce the structure of the ``vertical'' quantum cohomology generated by $\cO[\eta]$, including the quantum cohomology relation~(\ref{eq:math:qcci1}).

\subsubsection*{UV correlation functions}
It is straightforward to apply the methods of~\cite{Morrison:1994fr} 
to calculate the A-model genus-zero correlation functions of the operator corresponding to the hyperplane class,
represented by $\cO[\eta]$, in the UV phase.   In this large radius phase only Higgs vacua contribute, and we identify the operator ${\cal O}[\eta]$ with the GLSM field $\sigma$.

The anomalous chiral symmetry of the GLSM yields a selection rule for the non-zero correlation functions.
The non-vanishing A-model correlation functions of ${\cal O}[\eta]$ are of the form
\begin{align}
\langle {\cal O}[\eta]^{n-1 + m(n+1-d)}\rangle \: = \: A_m q^m~,
\end{align}
where $A_m$ is a $q$-independent constant.  Note that while we restrict to correlation functions that are polynomial in the $\cO[\eta]$, or in other words to $m$ in the range $-m \le (n-1)/(n+1-d)$, the selection rule does in principle allow for correlation functions with $m<0$.  However, each correlation function receives contributions from exactly one instanton number---$m$, and in fact $m\ge 0$ in the UV phase because for $m<0$ the instanton moduli space is empty.  Denoting the $m$-th instanton contribution as
\begin{align}
A_m \: = \: \langle {\cal O}[\eta]^{n-1 + m(n+1-d)}\rangle_{m}^M~,
\end{align}
we now use the restriction result of~\cite{Morrison:1994fr} to relate the correlation function in the A-model for $M$ to that for the A-model for $\mathbb{P}^{n}$:
\begin{align}
A_m \: = \: \langle {\cal O}[\eta]^{n-1 + m(n+1-d)}\rangle_{m}^M
\: = \: A_m \: = \:
 - \langle {\cal O}[\eta]^{n-1 + m(n+1-d)} (-d {\cal O}[\eta])^{1+m d}\rangle_{m}^{\P^{n}}~,
\end{align}
or
\begin{align}
A_m & = \: d (-d)^{dm} \langle {\cal O}[\eta]^{n+ m(n+1)}\rangle^{{\mathbb P}^{n}}_{m} = d (-d)^{dm}~.
\end{align}
Putting that together, we obtain the desired correlation functions:
\begin{align}
\label{eq:UVcorrelators}
\langle {\cal O}[\eta]^{n-1+m(n+1-d)} \rangle \: = \:  \begin{cases} 0 & m<0~, \\ 
                                                                                   d ( (-d)^d q)^m~ & m\ge 0~.
                                                                                   \end{cases}
\end{align}
We see that the computation recovers the quantum cohomology relation~(\ref{eq:math:qcci1}) specialized to the hypersurface case:
\begin{align}
{\cal O}[\eta]^{n} \: = \: (-d)^d q {\cal O}[\eta]^{d-1}~.
\end{align}

\subsubsection*{IR correlation functions}

By using the GLSM fields and localization, the correlation functions $\langle {\cal O}[\eta]^k\rangle$ can also be calculated in the IR by the methods of~\cite{Morrison:1994fr} for the Higgs contribution and those of~\cite{Melnikov:2006kb} for the Coulomb contribution.  
The Coulomb contribution takes the following form:
\begin{align}
\langle {\cal O}[\eta]^k\rangle_{\text{C}} & = \: \sum_{\text{vac}} H^{-1} \sigma^k~,
\end{align}
where the sum is over critical points of the effective twisted chiral superpotential for the $\sigma$ field.  In our case these are simply the solutions to 
\begin{align}
\label{eq:CoulVac}
\sigma^{n+1-d} & = (-d)^d q~,
\end{align}
while the measure factor is
\begin{align}
H \: = \: \left(\frac{n+1}{\sigma} + \frac{d^2}{-d\sigma} \right) \times \sigma^{n+1} \times \frac{1}{d\sigma}~. 
\end{align}
Let us review how the terms in the measure factor arise.  For a fixed Coulomb vacuum, the first factor comes from the integral over the fluctuations in the zero modes of the gauge sector.  Similarly, the second factor comes from the integral over the zero modes of the  A-twisted chiral multiplets $X_i$.  Finally, the last factor comes from the zero modes of A-twisted chiral multiplet $\Phi$.  Compared to~\cite{Melnikov:2006kb}, there is just one unusual feature here: the contribution from $\Phi$.  The reason for the modification is that $\Phi$ has vector R-charge $+2$.  This leads to a flip in the contribution from the ratio of the bose/fermi determinants, and there is a minus sign from rearranging the zero modes, as in~\cite{Morrison:1994fr}.

Performing the sum over the Coulomb vacua we then obtain the Coulomb contribution to the correlation function:
\begin{align}
\langle {\cal O}[\eta]^{n-1+m(n+1-d)}\rangle_{\text{C}} & = \: d ((-d)^d q)^m~.
\end{align}
While superficially agreeing with the UV computation when $m\ge 0$, the result is puzzling, since for $0<-m \le (n-1)/(n+1-d)$ there are non-zero contributions to correlation functions with $m<0$ yet non-negative exponent, i.e.
$\langle{\cal O}[\eta]^{d-2}\rangle_{\text{C}}$,  $\langle{\cal O}[\eta]^{2d-3-n}\rangle_{\text{C}}$, \ldots.  

Now we move onto the Higgs calculation, following~\cite{Morrison:1994fr}.  Here
\begin{align}
\langle {\cal O}[\eta]^{p}\rangle_{\text{H}}
 \: = \:
 \frac{1}{d} q^m \#\left( (-\delta_0^2) (\eta^{\ast})^{ p} \chi_{m}\right)_{{\cal M}_m}~,
\end{align}
where ${\cal M}_m$ is the instanton moduli space, $\eta^\ast$ is the pull-back of $\eta$ to ${\cal M}_m$, $\chi_{m}$ 
is the Euler class of the obstruction bundle, and $\delta_0 = (-d) \eta^\ast$.  
In the IR phase we find that ${\cal M}_{m}$ is empty for $m\ge 0$, while for $m<0$ ${\cal M}_{m} = {\mathbb P}^{-m d}$.  
The Euler class is given entirely by contributions from the $X_i$ and has the form
\begin{align}
\chi_{m} \: = \: \left( (\eta^{\ast})^{-m-1}\right)^{n+1}~.
\end{align}
Note the overall factor of $1/d$:  this comes from the unbroken ${\mathbb Z}_d$ gauge group.  Finally, the intersection on ${\cal M}_{m}$ is determined by
\begin{align}
\#(  ((-d) \eta^\ast)^{-m d} ) \: = \: 1~,
\end{align}
and putting all of that together we obtain that the Higgs branch makes non-zero contributions
\begin{align}
\langle {\cal O}[\eta]^{n-1+m(n+1-d)}\rangle_{\text{H}} \: = \: -d ((-d)^dq)^m~
\end{align}
for $m <0$.  Summing up the contributions we then obtain
\begin{align}
\langle {\cal O}[\eta]^{n-1+m(n+1-d)}\rangle 
&\: = \:
\langle {\cal O}[\eta]^{n-1+m(n+1-d)}\rangle_{\text{H}} +
\langle {\cal O}[\eta]^{n-1+m(n+1-d)}\rangle_{\text{C}}~\nonumber\\
&\: = \:  \begin{cases}  0 			& m <0~, \\
                                  d ((-d)^d q)^m ~,	& m\ge 0~.
          \end{cases}
\end{align}
This matches the UV result~(\ref{eq:UVcorrelators}), and it does so in a non-trivial fashion:  the Higgs sector makes contributions at instanton numbers where the UV computations has no contributions, while the Coulomb sector makes contributions at all instanton numbers.  Their sum, however, matches the UV Higgs computation, and this match provides a good check of our understanding of the physics, and in particular the decoupling of the Higgs and Coulomb degrees of freedom.

In principle it should be possible to extend these GLSM computations to include correlation functions with insertions of the $\cO[\alpha]$ operators.  Unfortunately, while the $\cO[\alpha]$ operators have a straightforward realization in the A-twisted non-linear sigma model, it is not obvious how to write these directly in terms of the GLSM fields.   This is a manifestation of a general problem in the GLSM description of geometries: a recent discussion may be found in~\cite{Adams:2023imc}.  So, while we can certainly count their multiplicities and organize them by their $q_V$ and $\rho$ charges, we cannot directly reproduce the remaining quantum cohomology relations by working in the UV phase beyond observing that the relations are consistent with the selection rules based on the symmetries.  We will see that the IR description gives a complementary perspective on these operators as arising from the twisted sector of the LGO sector.

The match of the UV and IR correlation functions, and their consistency with the quantum cohomology relation~(\ref{eq:math:qcci1}) supports the proposed decoupling between the Coulomb and Higgs sectors of the IR theory, but it only tests the quantum cohomology relations in the ``vertical'' column of the Hodge diamond.  We now turn to a more detailed study of the IR phase with the goal of describing an isomorphism between $\cH_{\text{uv}}$ and the A-model Hilbert space in the IR phase description.

\subsection{The Coulomb vacua in the IR phase}

As discussed above, we expect that the Coulomb and Higgs sectors decouple, and we can describe the operators and states in the two sectors separately.  In this section we will discuss the Coulomb sector, leaving the more involved Higgs phase to the next section.

The Coulomb vacua are labeled by expectation values of the field $\sigma$ determined by~(\ref{eq:CoulVac}).
The expectation value $\langle \sigma\rangle$ breaks the $\rho$ symmetry to $\Z_2$---the fermion number symmetry of the theory.\footnote{When working in a quantum field theory in infinite spatial volume we would have super-selection sectors labeled by the $\sigma$ expectation values.  In the finite-dimensional Hilbert space of the A-model we do not have such super-selection sectors, and instead the field $\sigma$ relates one vacuum to another.}
These $n+1-d$ massive vacua correspond to $n+1-d$ states in the A-model, which we denote by $|t;\text{C}\rangle$, with $t = 0,1,\ldots, n-d$, and the $\sigma$ field acts on these states by $\sigma |t;\text{C}\rangle = |t+1;\text{C}\rangle$ for $t = 0,1,\ldots,n-d-1$.  By taking suitable linear combinations we choose the state $|0;\text{C}\rangle$ to be invariant under the chiral $\rho$--symmetry, and the charges of the remaining states are determined by the action of $\sigma$.

These states have the quantum numbers that appear to match those of the UV phase states $|\eta^k;\text{uv}\ra$ for $k = 0,\ldots, n-d$, but we clearly see that to match the remaining states in $\cH_{\text{uv}}$ additional states are needed in the IR description---states that have the quantum numbers of $|\eta^k;\text{uv}\rangle$ with
\begin{align}
\label{eq:missingk}
n+1 -d \le k \le n~,
\end{align}
as well as states corresponding to the $|\alpha;\text{uv}\rangle$.
These additional states are provided by the Higgs vacua in the IR phase, and their physics is described by a Landau-Ginzburg orbifold.

\subsection{The LGO sector}

In the Higgs sector of the IR phase the chiral superfield field $\Phi$ acquires an expectation value which breaks the gauge symmetry to a subgroup $G=\Z_d$.  In the low energy limit the massive $\Phi$ and vector multiplet degrees decouple, 
leaving the low energy physics of a (2,2) supersymmetric Landau-Ginzburg orbifold (LGO),
where the superpotential $W$ is a degree
$d$ function of $n+1$ chiral superfields $X_0, \cdots, X_n$,
and the orbifold group is the unbroken gauge symmetry $G$.

Starting with the seminal work~\cite{Martinec:1988zu,Vafa:1988uu}, there is by now ample evidence that the theory has non-trivial IR dynamics of a (2,2) superconformal theory, with left and right central
charges given by
\begin{equation} \label{eq:c}
c \: = \: 
c_L \: = \: c_R \: = \: 3 \sum_i \left( 1 - 2 \theta_i \right)
\: = \: 3 (n+1) \left( \frac{d-2}{d} \right),
\end{equation}
where $\theta_i$ is the weight of $X_i$ chosen such that $W$ has
weight $1$.\footnote{The $\theta_i$ encode the R-charges of the chiral fields and should not be confused with the superspace coordinates $\theta^\pm$.}  For generic superpotential this is a non-trivial superconformal field theory (although it is a solvable Gepner-like theory when $W$ is taken to be Fermat).   We of course do not expect to be able to relate the dynamics of the full theory to the quantum geometry of the Fano hypersurface in the UV phase.  However, as we will see, the A-model topological sector of the LGO will provide exactly the states we need to match those of the UV A-model.  

We will compute the (a,c) ring of the (2,2) SCFT appearing as the IR limit
of the ${\mathbb Z}_d$ orbifold of the Landau-Ginzburg theory above,
as those states are the ones that contribute to the A model.
Our analysis will follow \cite{Intriligator:1990ua}, to which we
refer the reader for additional details.

We should note one important difference between this LGO and 
those that are typically 
discussed in the context of the Calabi-Yau/LGO correspondence,
regarding the non-local spectral flow fields ${\cal U}_{\alpha_L,\alpha_R}$ that shift the charges of operators according to  
\begin{equation}
(q_L,q_R) \: \mapsto \: (q_L + \alpha_L \, c/3, q_R + \alpha_R \, c/3).
\end{equation}
In that perhaps more familiar setting of the Calabi-Yau/LGO correspondence,
the SCFT has integral $q_L$ and $q_R$ charges after the orbifold projection, 
and thus chiral spectral flow operators ${\cal U}_{1/2,0}$, ${\cal U}_{0,1/2}$ 
that can be used to relate chiral R and NS sectors,
and construct the R-NS and NS-R sectors and a consistent string vacuum, 
as discussed in~\cite{Vafa:1989xc}.  
In our context the projection will not lead to integral $q_L$ and $q_R$ charges 
essentially because $c/3 \not \in \Z$ in general, and hence we do not have chiral
spectral flow operators.  
Nevertheless, the twisting procedure for the $\Z_d$ orbifold 
can still be formally implemented 
using the 
spectral flow operator ${\cal U}_{-1/2,1/2}$ and its inverse.
For the particular orbifold we consider this leads to an important isomorphism between the unprojected RR states and the unprojected (a,c) ring states:
if we let ${\cal H}_{ac}^{\rm unproj}(r)$ denote the (a,c) ring in
the $r$ twisted sector before projecting to $G$-invariant states,
and ${\cal H}_{RR}^{\rm unproj}(r)$ denote the RR sector states
in the $r$ twisted sector before projecting to $G$-invariant states,
then
\begin{equation}
{\cal U}_{1/2,-1/2}: \: {\cal H}_{ac}^{\rm unproj}(r) \: 
\stackrel{\sim}{\longrightarrow} \:
{\cal H}_{RR}^{\rm unproj}(r-1).
\end{equation}
As discussed in~\cite{Intriligator:1990ua}, the isomorphism of the unprojected (a,c) states in a sector twisted by an element $h$ and that of the RR states in a sector twisted by an element $hj^{-1}$ holds in general LGOs.  The special simplifying feature in our case, where the orbifold is by the $\mathbb{Z}_d$ symmetry generated by $j$ is that ${\cal U}_{1/2,-1/2}$ is an operator in the projected theory and yields an isomorphism for the projected states as well.

With this in mind, the algorithm we will follow to compute the
twisted-sector (a,c) states is to first
classify the unprojected RR twisted sector states, then apply
spectral flow ${\cal U}_{-1/2,1/2}$ to get the unprojected (a,c) ring states,
and finally take $G$-invariants to get the desired A model contributions.

In our case the orbifold group $G= {\mathbb Z}_d$ has generator $j$, which
acts on the chiral superfields $X_i$ by
\begin{align}
j \cdot X_i & =  \exp\left(2 \pi i \theta_i \right) \, X_i~,&
\theta_i & = \frac{1}{d}~,
\end{align}
where the second equation is valid for a degree $d$ hypersurface in ${\mathbb P}^n$.

Let $|r; {\rm RR} \rangle$ denote the RR sector vacuum in the $r$ twisted
sector with $0\le r <d$.  The (unprojected) states in the $r$ twisted RR sector are
of the form
\begin{equation}
f_p(x) |r; {\rm RR} \rangle
\end{equation}
for $f_p(x)$ a homogeneous polynomial of degree $p$ in the $x_i$---the zero modes of the scalar fields in the multiplets $X_i$.  If $r > 0$, then because of the form of the $G$ action,
all of the $x_i$ have nonzero moding, and so the only possible ground states have $p=0$.  On the other hand, for $r=0$ every $x_i$ has a zero mode, and the ideal
\begin{align}
\left( dW \right) = \left(~ \frac{\partial W}{\partial x_0}, \frac{\partial W}{\partial x_1},\ldots, \frac{\partial W}{\partial x_n} ~\right) \in \C[x_0,x_1,\ldots,x_n]~ 
\end{align}
annihilates all of the states, so that we can take $f_p(x) \in \cR$, where $\cR$ is the Jacobian ideal $\cR = \C[x_0,x_1,\ldots,x_n]/\left( dW\right)$.  When we write $f_p(x)$ it will be with this understanding.

Unprojected (a,c) states in the $r$ twisted sector are of the form
\begin{equation}
f_p(x) |r;\text{ac}\rangle = {\cal U}_{-1/2,1/2} \, f_p(x) | r-1; {\rm RR} \rangle.
\end{equation}
To obtain the projected states we need the action of $j$ in the twisted sectors.  This is worked out in~\cite{Intriligator:1990ua}, and for our group action the result simplifies to
\begin{equation}
j \cdot  f_p(x) | r; {\rm ac} \rangle
\: = \: \exp\left[ 2 \pi i \left( \theta^j_T(r-1) + \frac{p}{d} \right)
\right] \,
 f_p(x) | r; {\rm ac} \rangle~,
\end{equation}
where $\theta^j_T(r-1)$ denotes the contribution from the action of $j$ on the
Landau-Ginzburg fields $X_i$ which are invariant under $j^{r-1}$.

We can summarize results as follows:
\begin{itemize}
\item $r=0$:  There is one untwisted (a,c) state that survives the orbifold.
\item $r=1$:  Here, 
\begin{equation}
\theta^j_T(r-1) \: = \: \sum_i \theta_i \: = \: \frac{n+1}{d}
\end{equation}
(each of $n+1$ Landau-Ginzburg fields $x_i$ contributes $\theta_i = 1/d$).
Therefore, we project onto states for which
\begin{equation}  \label{eq:constr-p}
 p + n + 1  \equiv 0 \mod d~.
\end{equation}
\item $1 < r < d$:  Here, only states with $p=0$ contribute, and since no Landau-Ginzburg fields are invariant under
$j^{r-1}$, $\theta^j_T(r-1) = 0$, and each state $|r;\text{ac}\ra$ is $G$--invariant.
\end{itemize}

The Landau-Ginzburg orbifold has a $U(1)_L\times U(1)_R$ symmetry, and 
we classify states according to their charges $q_L$, $q_R$.  These are determined
as follows.  First, the R-charges of the RR vacuum $|r, {\rm RR} \rangle$ are given by
\cite{Wen:1985qj}
\begin{eqnarray}
q_L & = & + \sum_{i \not\in {\cal T}_r} \left( \theta^r_i - [ \theta^r_i ]
- \frac{1}{2} \right)
 \: + \:
\sum_{i \in {\cal T}_r} \left( \theta_i - \frac{1}{2} \right),
\\
q_R & = & - \sum_{i \not\in {\cal T}_r} \left( \theta^r_i - [ \theta^r_i ]
- \frac{1}{2} \right)
 \: + \:
\sum_{i \in {\cal T}_r} \left( \theta_i - \frac{1}{2} \right),
\end{eqnarray}
where here $\theta^r_i = r/d$ for all $i$, and
\begin{equation}
{\cal T}_r \: = \: \{ i \in \{1, \cdots, n+1\} \, | \,
\theta_i^r \in {\mathbb Z} \}.
\end{equation}
We can simplify this expression for charges of RR vacua
by distinguishing two cases:
\begin{itemize}
\item $r=0$:  Here, the RR vacuum $|0, {\rm RR} \rangle$ has charges
\begin{equation}
q_L \: = \: + q_R \: = \:
(n+1) \left( \frac{1}{d} - \frac{1}{2} \right)
\: = \: - \frac{c}{6}.
\end{equation}
\item $0 < r < d$:  Here,
\begin{equation}
q_L \: = \: - q_R \: = \: (n+1) \left( \frac{r}{d} - \frac{1}{2} \right).
\end{equation}
\end{itemize}

The RR state $f_p(x)|r, {\rm RR} \rangle$ has $q_{L,R}$ differing from those
of $|r, {\rm RR} \rangle$ by the addition of $p/d$, the R-charge of $f_p(x)$.
The spectral flow operator ${\cal U}_{-1/2,+1/2}$ has
\begin{equation}
(q_L, q_R) \: = \:  \left(
- \frac{c}{6}, + \frac{c}{6} \right)~,\qquad\qquad \frac{c}{3} = (n+1) \left( 1 -\frac{d}{2}\right)~,
\end{equation}
and from this we deduce that the $r$-twisted (a,c) state
\begin{equation}
f_p(x) |r;\text{ac}\rangle = {\cal U}_{-1/2,1/2} \, f_p(x) | r-1; {\rm RR} \rangle
\end{equation}
has R-charges given as follows:
\begin{itemize}
\item $r=0$:  Here, the corresponding RR state is in sector $d-1$, and we
find
\begin{equation}
q_L \: = \: + q_R \: = \: \frac{p}{d}.
\end{equation}
\item $r=1$:
\begin{equation}
q_L \: = \: \frac{p}{d} - \frac{c}{3} \: = \:
\frac{p}{d} - (n+1) \left( \frac{d-2}{d} \right),
 \: \: \:
q_R \: = \: \frac{p}{d}.
\end{equation}
\item $1 < r < d$:
\begin{equation}
q_L \: = \: \frac{p}{d} + (n+1) \left( \frac{r}{d} - 1 \right),
\: \: \:
q_R \: = \: \frac{p}{d} - (n+1) \left( \frac{r}{d} - 1\right).
\end{equation}
\end{itemize}
Applying the algorithm we then obtain the Hilbert spaces of the projected (a,c) states $\cH_{\text{ac}}(r)$ in the $r$-th twisted sector:
\begin{align}
\cH_{\text{ac}} (0) & = \text{Span} \{ |0;\text{ac}\ra \}~, \nonumber\\
\cH_{\text{ac}}(1) & = \text{Span}\{ f_p|1;\text{ac}\ra ~~|~~ f_p \in \cR~, p +n+1 \equiv 0 \mod d\}~, 
\end{align}
and for $1<r <d$
\begin{align}
\cH_{\text{ac}}(r) & = \text{Span} \{ |r;\text{ac}\ra\}~.
\end{align}

\subsubsection*{Matching the UV and IR symmetries}
We now have both UV and IR descriptions of the A-model states space.  We will use symmetries preserved along the RG flow of the physical theory to match the two descriptions.  Recall that we identified a continuous non-chiral vector symmetry, as well as a discrete chiral R-symmetry.  

The vector symmetry with charges $q_V$ is present throughout the RG flow, and we know precisely how it acts on the GLSM fields in either phase.  We therefore expect that the $q_V = q_R + q_L$ in the LGO theory as well.  With this identification made we see that the only states in the LGO that can match the middle cohomology states $|\alpha;\text{uv}\rangle$ are those from the $r=1$ twisted sector.

In the UV phase we also defined the discrete symmetry with generator $\rho$. In the IR its fate is complicated:  it is spontaneously broken to a $\mathbb{Z}_2$ in the Coulomb sector, and while it remains unbroken in the LGO vacuum, it is not a priori clear how to identify it with symmetries in that vacuum.  The relevant symmetries in the LGO vacuum are the chiral R-symmetry with charges $q_R- q_L$ and the quantum symmetry of the orbifold, which assigns a charge $\exp\{2\pi i r/d\}$ to the $r$-th twisted sector.   To make the identification, we note that the $\rho$-charge of $\cO[\alpha]$ for $\alpha \in H^{n-1}(X)$ is $\zeta^{n-1}$, and as we just observed, we expect these states to map to the first twisted sector of the LGO.  Moreover, for all LGO states $q_R - q_L + r c/3 \in \mathbb{Z}$.  Putting these two observations together, we arrive at the following proposal for the action of $\rho$ on the LGO states:
\begin{align} \label{eq:rho-action}
\rho \cdot f_p(x) | r; {\rm ac} \rangle &=\exp\left[ \frac{\pi i}{n+1-d} \left( q_R - q_L + 
\frac{2 (n+1-d) r}{d} \right) \right] f_p(x) | r; {\rm ac}\rangle~ \nonumber\\
& = \zeta^{q_R-q_L} e^{2\pi i r/d} f_p(x) | r; {\rm ac}\rangle~.
\end{align}
It is easy to check that this generates a $\mathbb{Z}_{2(n+1-d)}$ action on all of the LGO states, and we will see that it leads to a nice matching between the UV and IR descriptions of the A-model's Hilbert space.

\subsubsection*{Multiplicities in the $r=1$ twisted sector}
As we just saw, the description of the LGO states is simple, and the only non-universal feature concerns the multiplicities of the states in the $r=1$ twisted sector.  While calculating these multiplicities in particular examples is easily accomplished, the explicit enumeration is awkward.  Fortunately, there is an elegant generating function that encodes all of the multiplicities.

Let us start by defining a generating function that counts the projected (a,c) states in the first twisted sector, i.e.
\begin{align}
F_{n+1}(t,\tb) = \Tr_{\cH_{\text{ac}}(1)} t^{d q_{L}} \tb^{d q_{R}} = \sum_{q_{L},q_{R}} \mu_{dq_{L},dq_{R}}  t^{d q_{L}} \tb^{d q_{R}}~.
\end{align}
Equivalently we can write this as a sum over the unprojected states with an insertion of a projector onto states where\footnote{This form of the projection is equivalent to the one given above for $f_p|1;\text{ac}\ra$ as $p+n+1 = 0 \mod d$, but it is the more convenient form for the generating function manipulations that will follow.}
\begin{align}
d q_{L} -(n+1) = 0 \mod d~.
\end{align}
This leads to
\begin{align}
F_{n+1}(t,\tb) = \frac{1}{d} \sum_{k=0}^{d-1} \Tr_{\cH^{\text{unproj}}_{\text{ac}}(1)} \left[ t^{d q_{L}} \tb^{d q_{R}} \xi^{k(dq_{L} - (n+1))}\right]~,
\end{align}
where the trace is now over the unprojected states, and $\xi = e^{2\pi i/d}$.  We know the explicit generating function for the unprojected states~\cite{Intriligator:1990ua,MR2896292}:
\begin{align}
M(t,\tb) =\Tr_{\cH^{\text{unproj}}_{\text{ac}}(1)} t^{d q_{L}} \tb^{d q_{R}}  = t^{-dc/3} \Tr_{\cH^{\text{unproj}}_{\text{cc}}(1)} t^{dq_{L}} \tb^{dq_{R}} = t^{-dc/3} \left( \frac{ (1-(t\tb)^{d-1})}{1-t\tb}\right)^{n+1}~,
\end{align}
so we have
\begin{align}
F_{n+1}(t,\tb) & = \frac{1}{d} \sum_{k=0}^{d-1} \xi^{-k(n+1)} M(t\xi^k,\tb)~.
\end{align}
Simplifying further,
\begin{align}
F_{n+1}(t,\tb) 
=\frac{1}{d} \sum_{k=0}^{d-1} \left( \frac{\xi^{k} t^{2-d} - t \tb^{d-1}}{1- \xi^k t\tb}\right)^{n+1} ~.
\end{align}
For any fixed $n$ this gives a nice way of listing the projected states, but it is awkward to implement the projection.  Much the same awkwardness arises when trying to calculate the middle cohomology Hodge numbers $h^{p,q}(X)$ reviewed above, and this inspires us to instead consider a generating function obtained by summing over all possible $n$:
\begin{align}
G(s,t) = \sum_{n=0}^{\infty} s^{n} F_{n+2}(t,t)~.
\end{align}
Notice that we also set $t=\tb$, so that the coefficient of $(t^{d})^a$ in the expansion counts the multiplicity of states with a fixed value of  the $q_V$ charge $q_V = q_L+q_R=a$.

After performing the geometric sum on $n$ we can simplify $G(s,t)$ and carry out the sum that implements the projection through repeated use of the identity
\begin{align}
\sum_{l=0}^{d-1} a^{d-1-l} z^l = \frac{a^d - z^d}{a-z}~,
\end{align}
which allows us to shift all of the $\xi$--dependence into the numerator of the rational function.  The result is
\begin{align}
G(s,t) & = \frac{ ( 1+ st^{-d})^{d-1} - (1+st^d)^{d-1}}{(1+st^d)^dst^{-d}-(1+st^{-d})^d st^d}~.
\end{align}
We now compare this to the generating function $H(x,y)$ for the primitive Hodge numbers from~(\ref{eq:gen-fn-hodge}) above and observe that
\begin{align}
H(s/t^d,s t^d) = G(s,t)~.
\end{align}
But now since
\begin{align}
H(s/t^d,s t^d) = \sum_{k,l} \mu_{k,l} s^{k+l} t^{d(l-k)}~,
\end{align}
we see that the multiplicities of the $r=1$ twisted sector states in the LGO precisely match the primitive Hodge numbers once we make the identification $n= k+l+1$ and $q_L+q_R = l-k$.

\subsection{Matching the UV and IR descriptions}
\label{sect:matching}

Having analyzed the A-model's Hilbert space of states from both the UV and IR points of view, we will now use the identification of symmetries to match, as far as possible, the two presentations.  To  do so, we first review our findings, organizing the states by their charges.

In the UV description we have the presentation of the A-model Hilbert space as 
\begin{align}
\cH_{\text{uv}} \simeq \cH_{\text{uv}}^{\text{vert}} \oplus \textstyle\bigoplus_{k+l=n-1} \cH_{\text{uv}}^{k,l}~,
\end{align}
with
\begin{align}
\cH^{\text{vert}}_{\text{uv}} &=  \text{Span}\{ \cO[\eta]^k |\Omega;\text{uv}\ra,~~k=0,\ldots,n-1\}~,
\nonumber\\[2mm]
\cH_{\text{uv}}^{k,l}  &= \text{Span} \{\cO[\alpha_{k,l}]|\Omega;\text{uv}\ra~,\alpha \in H^{k,l}_{\text{prim}}(X)\}~,
\end{align}
and we characterize the states according to their symmetry charges as
\begin{align}
\text{state} &&  \cO[\eta]^k |\Omega;\text{uv}\ra 	&&   \cO[\alpha_{k,l}]|\Omega;\text{uv}\ra^{\oplus \mu_{k,l}} \nonumber\\
q_V &&  0	~~~~~~						&& l-k ~~~~~~~~~\nonumber\\
\rho &&  \zeta^{2k}~~~~					&& \zeta^{l+k}~~~~~~~~~.
\end{align}

The observables have a more complicated description in the IR variables.   The Hilbert space is a direct sum of the Coulomb and LGO factors:
\begin{align}
\cH_{\text{ir}} = \cH_{\text{C}} \oplus \cH_{\text{LGO}}~,
\end{align}
and the latter has a further decomposition into the orbifold sectors:
\begin{align}
\cH_{\text{LGO}} = \cH_{\text{ac}}(1)\oplus \textstyle\bigoplus_{r=2}^{d} \cH_{\text{ac}}(r)~.
\end{align}
Note that we included the untwisted sector as $r=d$.  While the $r>1$ spaces are all one-dimensional and carry $q_V =0$, $\cH_{\text{ac}}(1)$ can be graded further by the $q_V$ charges:
\begin{align}
\cH_{\text{ac}}(1) = \textstyle\bigoplus_{q_V} \cH_{\text{ac}}^{q_V}(1)~.
\end{align}
We determined the action of $\rho$ in the LGO sector by requiring that the states in $\cH_{\text{ac}}(1)$ carry $\rho$--charge $\zeta^{n-1}$, and we showed above that 
\begin{align}
\dim \cH_{\text{ac}}^{l-k}(1) = \dim H^{k,l}_{\text{prim}}(X)~,
\end{align}
and we therefore propose that there is an isomorphism $\cH_{\text{uv}}^{k,l} \simeq \cH_{\text{ac}}^{l-k}(1)$.\footnote{
Note that we defined $\cH_{\text{uv}}^{k,l}$ as the primitive Dolbeault cohomology of $X$.  It is possible to grade these spaces further by the large permutation symmetry enjoyed by the A-model and thereby refine the isomorphism.}

The remaining states in the IR description consist of the $n+1-d$ Coulomb states $|t;\text{C}\ra$ and the $d-1$ LGO states from the untwisted sector $|0;\text{ac}\ra$ and the higher twisted sectors $|r;\text{ac}\ra$ with $1<r<d$.  All of these states have $q_V = 0$, while their $\rho$--charges are as follows:
\begin{align}
\text{state} && |0;\text{C}\ra   && \cdots && |t;\text{C}\ra && \cdots && |n-d;\text{C}\ra \nonumber\\
\rho
       &&   \zeta^0 ~~        && \cdots   && \zeta^{2t}   ~~  && \cdots && \zeta^{2(n-d)}~~~,
\end{align}
and 
\begin{align}
\text{state} && |j^d;\text{ac}\ra   && \cdots && |j^{d-r};\text{ac}\ra && \cdots && |j^{2};\text{ac}\ra ~~~\nonumber\\
{\rho} 
       &&   \zeta^0 ~~        && \cdots   && \zeta^{2r}   ~~  && \cdots && ~~\zeta^{2(d-2)}~.
\end{align}
We would like to match these $n$ states to those in 
$\cH_{\text{uv}}^{\text{vert}}$. The dimensions match, but we do not have
enough symmetry to form a unique one-to-one correspondence between states and 
cohomology.  Instead, we have at least a two-fold ambiguity in
sectors with $\rho$ charge $\zeta^{2r}$ for $r =0,\ldots,d-2$,
in both the IR description, as well as
in the UV description, since both $\cO[\eta^{r}]$ and $\cO[\eta^{n+1-d+r}]$ have $\rho$--charge $\zeta^{2r}$.  
There will be yet further ambiguity in the spectrum if\footnote{
We observe that this is the same constraint~(\ref{eq:n-constr}) that
appears in \cite{beauville}, albeit the origin here is completely different.
} $n<2d-2$.

To proceed with the identification we observe $\cH_{\text{uv}}^{\text{vert}}$ is generated by powers of a single operator $\cO[\eta]$ acting on a vacuum state $|\Omega;\text{uv}\ra$ that we associate to the identity operator.  Similarly, the Coulomb Hilbert space is generated by the action of $\sigma$:
\begin{align}
\cH_{\text{C}}& \simeq \text{Span}\{ \sigma^k | 0;\text{C}\ra,~~ k = 0,\ldots, n-d\}~.
\end{align}
In the LGO sector we have the state $|0;\text{ac}\ra$, and by the state-operator correspondence we can find an operator $\Psi$ of $\rho$--charge $\zeta^2$ and quantum symmetry charge $e^{-2\pi i/d}$ such that
\begin{align}
|d-1;\text{ac}\ra = \Psi |0;\text{ac}\ra~.
\end{align}
Taking further powers we obtain states $\Psi^k|0;\text{ac}\ra$  with the same quantum numbers as $|d-k;\text{ac}\ra$ for $0 \le k \le d-2$, while for $\Psi^{d-1} |0;\text{ac}\ra$ we obtain a state with the quantum symmetry charge of the first twisted sector but $\rho$--charge $\zeta^{2(d-1)}$.  This is is inconsistent with the $\rho$--charge in the first twisted sector, which is given by $\zeta^{n-1}$ unless
\begin{align}
2(d-1) = n-1 \mod (n+1-d)~.
\end{align} 
But this is impossible, since it is equivalent to $d = m(n+1-d)$ for some positive integer $m$, and that is inconsistent with our basic assumption $d < n+1$.  Thus it must be $\Psi^{d-1} = 0$.  We will assume that $\Psi^{k} \neq 0$ for $k <d-1$, so that $\Psi$ generates all of the higher twisted states:
\begin{align}
\oplus_{r=2}^{d} \cH_{\text{ac}}(r) = \text{Span} \{ \Psi^k|0;\text{ac}\ra ~~| ~~0\le k \le d-2\}~.
\end{align}
Having made these identifications we now proceed to describe our proposal for the isomorphism
\begin{align}
\cH_{\text{uv}}^{\text{vert}} \simeq \cH_\text{C}\oplus \oplus_{r=2}^{d} \cH_{\text{ac}}(r)~.
\end{align}
We suppose that the ground state can be written as
\begin{align}
|\Omega;\text{uv}\ra = |0;\text{C}\ra + |0;\text{ac}\ra~,
\end{align}
and identify
\begin{align}  \label{eq:hyperplane}
\cO[\eta]  = \sigma + x\Psi~,
\end{align}
where $x$ is a parameter that we will constrain further momentarily.  Our fields have the following properties:
\begin{align}
\sigma^{n+1 -d } & = (-d)^d q~,&
\Psi^{d-1} & = 0~, &
\sigma \Psi & = 0~,&
\sigma |0;\text{ac}\ra & = 0~,&
\Psi |0;\text{C}\ra & =0~,
\end{align}
where the last three relations follow from the decoupling between the Higgs and Coulomb sectors.  
We now calculate, for $d>1$,
\begin{align}   \label{eq:final:c1}
\cO[\eta]^{n} | \Omega;\text{uv}\ra =  \sigma^{n} |0;\text{C}\ra = \sigma^{n+1-d} \sigma^{d-1} |\Omega;\text{uv}\ra
= (-d)^d q \cO[\eta]^{d-1} | \Omega;\text{uv}\ra~,
\end{align}
and this is one of the UV quantum cohomology relation~(\ref{eq:phys:qcci1}).

Next, we return to the states in $\cH_{\text{ac}}(1)$, which we already identified with $H^{n-1}_{\text{prim}}(X)$.  With this identification the state-operator correspondence implies that for each $\alpha \in H^{n-1}_{\text{prim}}(X)$ there is a LGO field $\Xi[\alpha]$ of $\rho$-charge $\zeta^{n-1}$ such that
\begin{align}
\cO[\alpha] & = \Xi[\alpha] | 0;\text{ac}\ra~.
\end{align}
Applying the selection rules as before we conclude that
\begin{align} 
\Xi[\alpha] \Psi = 0~.
\end{align} 
Furthermore, the Coulomb/Higgs decoupling implies
\begin{align}  \label{eq:final:root}
 \Xi[\alpha] \sigma &= 0\qquad\qquad &\text{and}&& \qquad\qquad \Xi[\alpha] |0;\text{C}\ra &= 0~.
 \end{align}  
This is another one of the UV quantum cohomology relations~(\ref{eq:phys:qcci1}).

 On the other hand, $\Xi[\alpha]\Xi[\beta]$ carries quantum symmetry charge $e^{4\pi i /d}$ and $\rho$--charge $\zeta^{2(n-1)} = \zeta^{2(d-2)}$, so for $d>2$ it must be that
\begin{align}
\Xi[\alpha]\Xi[\beta] |\Omega;\text{uv}\ra = C(\alpha,\beta) \Psi^{d-2} |\Omega;\text{uv}\ra~.
\end{align}
But now we observe, with $d>2$,
\begin{align}
\cO[\eta]^{n-1} |\Omega;\text{uv}\ra  &= (\sigma + x \Psi)^{n-1} |\Omega;\text{uv}\ra = \sigma^{n-1} |\Omega;\text{uv}\ra= (-d)^d q \sigma^{d-2} | \Omega;\text{uv}\ra~, \nonumber\\
\cO[\eta]^{d-2} |\Omega;\text{uv}\ra &= \left(\sigma^{d-2} + x^{d-2} \Psi^{d-2} \right) |\Omega;\text{uv}\ra~,
\end{align}
and putting these two statements together,
\begin{align}
\left(\cO[\eta]^{n-1} - (-d)^d q \cO[\eta]^{d-2} \right) |\Omega;\text{uv}\ra= -(-d)^d q x^{d-2} \Psi^{d-2} |\Omega;\text{uv}\ra~.
\end{align}
Eliminating $\Psi^{d-2}$, we therefore obtain
\begin{align}  \label{eq:final:lg1}
\Xi[\alpha] \Xi[\beta] |\Omega;\text{uv}\ra = \frac{C(\alpha,\beta)}{- (-d)^d qx^{d-2}}  \left(\cO[\eta]^{n-1} - (-d)^d q \cO[\eta]^{d-2} \right) |\Omega;\text{uv}\ra~.
\end{align}
This equation nearly corresponds 
to one of the desired quantum cohomology relations~(\ref{eq:phys:qcci2}).  
To complete the match,
we need to determine $C(\alpha,\beta)$ and the constant\footnote{
Recall that the constant $x$ determines the relative weight of $\sigma$ and LGO states
in the physical operator ${\cal O}[\eta]$ corresponding to the
hyperplane class, see equation~(\ref{eq:hyperplane}).
} $x$.
Now, on general principles, we strongly expect $C(\alpha,\beta)$ is
proportional to the product $(\alpha|\beta)$, so the only remaining
unknown is the overall proportionality constant, which can be absorbed
into field redefinitions.
Without loss of generality, we can fix that overall proportionality constant to
match the conventions of~(\ref{eq:phys:qcci2})  
by requiring
\begin{equation}
\frac{ C(\alpha,\beta) }{
- (-d)^d qx^{d-2}}
\: = \: \frac{ ( \alpha | \beta ) }{d}.
\end{equation}
For example, if $C(\alpha,\beta) = (\alpha | \beta)$, then
then this will reproduce the quantum cohomology relation~(\ref{eq:phys:qcci2}) obtained in \cite{beauville} provided that
\begin{equation}
- (-d)^d q x^{d-2} \: = \: d,
\end{equation}
or equivalently
\begin{equation}
 x  = -\frac{1}{d} (-dq)^{\frac{1}{d-2}}~.
\end{equation}
It would be illuminating to obtain the pairing $C(\alpha,\beta)$ 
from a direct LGO computation. 

To summarize, we proposed a detailed isomorphism $\cH_{\text{uv}} \simeq \cH_{\text{ir}}$ largely determined by symmetries, and we explained how the quantum cohomology structure on $\cH_{\text{uv}}$ can be understood in the IR phase as arising from selection rules, the decoupling of the Coulomb and Higgs sectors, and the structure of the Coulomb vacua.  Our proposal passes a number of consistency checks and relies on the identifications
\begin{align}
\label{eq:UVIRmap}
|\Omega;\text{uv}\ra & = |0;\text{C}\ra + |0;\text{ac}\ra~, \nonumber\\
\cO[\eta] & = \sigma - \frac{1}{d} (-dq)^{\frac{1}{d-2}} \Psi~, \nonumber\\
\cO[\alpha] & = \Xi[\alpha]~.
\end{align}
We have recovered the predicted
OPE relations in the mixed Higgs/Coulomb states~(\ref{eq:phys:qcci1}), 
(\ref{eq:phys:qcci2}),
namely
\begin{equation}  %\label{eq:phys:qcci1}
(\cO[\eta])^{n+1-m} = q \left( \prod_{a=1}^m \left(-d_a \right)^{d_a} \right)
(\cO[\eta])^{\sum_a (d_a - 1)}~, \: \: \: 
\cO[\eta] \cdot \cO[\alpha] = 0~, 
\end{equation}
\begin{equation}  %\label{eq:phys:qcci2}
\cO[\alpha] \cdot \cO[\beta] = (\alpha | \beta) \frac{1}{d}
\left( (\cO[\eta])^{n-m} - q \left( \prod_{a=1}^m  \left( - d_a \right)^{d_a}  \right)
(\cO[\eta])^{\sum_a (d_a - 1) - 1}  \right),
\end{equation}
corresponding to the product structure in the cohomology ring,
above as equations~(\ref{eq:final:c1}), (\ref{eq:final:root}),
and (\ref{eq:final:lg1}).
Furthermore, we note that the physical origin as Coulomb/Higgs
branch states defines a distinction in the cohomology that is different
from the role played by primitive cohomology.

\section{Examples}
\label{sect:exs}

Next, we will compute the predicted cohomology ring explicitly in
a number of examples.  We will verify in each case that the
predictions match known mathematics results.  For degrees $d \geq 3$,
to compute the Landau-Ginzburg states, we will specialize to Fermat (diagonal)
hypersurfaces.  Since the A-model is independent of the precise 
form of the chiral superpotential this choice is just a matter of convenience.

\subsection{Example: hyperplanes}
\label{sect:exs:deg1}

In this section we will consider the GLSM for a hyperplane,
namely ${\mathbb P}^n[1]$.
The case of hyperplanes is particularly simple, as
${\mathbb P}^n[1] = {\mathbb P}^{n-1}$.  We shall quickly walk through
the details to check that this special case is correctly reproduced.
(This analysis has also appeared in e.g.~\cite[section 5.2]{Gu:2020zpg}.)

From equation~(\ref{eq:midcoh:dim}),
\begin{equation}
D_p(n,1) \: = \: 0.
\end{equation}
Expanding the generating function~(\ref{eq:gen-fn-hodge}), we have
\begin{eqnarray}
\sum_{k,\ell} \left( h^{k,\ell} - \delta_{k,\ell} \right)
x^k y^{\ell}
& = &
0.
\end{eqnarray}
This means that the Dolbeault cohomology of ${\mathbb P}^n[1]$
is completely diagonal ($h^{p,q} = \delta_{p,q}$),
as expected for ${\mathbb P}^{n-1}$.

In physics, the Coulomb branch relation~(\ref{eq:qc:ci}) becomes
\begin{equation}
\sigma^{n+1-1} \: = \: - q \sigma^0,
\end{equation}
or more simply, $\sigma^n = -q$.  Hence, the $\sigma$ fields describe Dolbeault
cohomology $H^{k,k}$ for $k < n$, which matches the quantum cohomology ring
of ${\mathbb P}^{n-1}$, with no need for any additional contributions.

In this special case, the Landau-Ginzburg model has a linear superpotential,
which does not have any supersymmetric vacua (as $dW \neq 0$), hence there
is no contribution from the Landau-Ginzburg model, consistent with the
Coulomb branch computation.

\subsection{Example: quadric hypersurfaces}

Next, we turn to degree two hypersurfaces in ${\mathbb P}^n$.
(This case was also discussed in e.g.~\cite[section 5.3]{Gu:2020zpg};
we include this case for completeness.)
We assume the hypersurfaces are Fano.

First, we discuss the mathematics.
From equation~(\ref{eq:midcoh:dim}),
\begin{equation}
D_p(n,2) \: = \: \frac{1 + (-)^n}{2} + (-)^{n-1}
\: = \: 
\left\{ \begin{array}{cl}
0 & n \: {\rm even}, \\
1 & n \: {\rm odd}.
\end{array}\right.
\end{equation}
Expanding the generating function~(\ref{eq:gen-fn-hodge}), we have
\begin{eqnarray}
\sum_{k,\ell} \left( h^{k,\ell} - \delta_{k,\ell} \right)
x^k y^{\ell}
& = &
\frac{
(1+y) - (1+x)
}{
(1+x)^2 y - (1+y)^2 x
},
\\
& = &
\frac{ y-x }{ (y-x) - xy(y-x) } ,
\\
& = &
\frac{1}{1-xy} \: = \: \sum_{k=0}^{\infty} (xy)^k.
\end{eqnarray}
This is interpreted to mean
\begin{equation}
h^{n-1}\left( {\mathbb P}^n[2] \right)
\: = \: \left\{ \begin{array}{cl}
1 & n-1 \: {\rm odd}, \\
2 & n-1 \: {\rm even},
\end{array} \right.
\end{equation}
where the extra contribution for $n-1$ even corresponds to
monomials $x^{n-1} y^{n-1}$.

Mathematically,
in the case that $n+1$ is odd, the entire cohomology
ring is a restriction of the cohomology ring of
${\mathbb P}^n$ to the hypersurface, specifically,
\begin{equation}
{\mathbb C}[x] / (x^n - q),
\end{equation}
and in this case, we will see that it can be completely
understood from the Coulomb branch.
The quantum cohomology ring of ${\mathbb P}^n[2]$ for $n+1$ even
can be presented as
\begin{equation}  \label{eq:qh:evendiml-quadric}
{\mathbb C}[y,t,q] / \langle
y^{2k+1} - 4qy, \: \: \:
yt = 0, \: \: \:
t^2 = (-1)^k( y^{2k} - 4q )
\rangle
\end{equation}
where $n = 2k+1$, $y=1-x$.

Now, let us turn to the corresponding physics.
From the Coulomb branch, there are $\sigma$ fields, obeying the
relation~(\ref{eq:qc:ci})
\begin{equation}
\sigma^{n+1-1}  \: = \: q (-2)^2 \sigma,
\end{equation}
or equivalently for our purposes
\begin{equation}
\sigma^{n-1} \: = \:  (-2)^2 q.
\end{equation}
The Coulomb branch contributes $n-1$ states 
to the
Dolbeault cohomology groups $H^{k,k}$, of which there are a total
of $n$ (corresponding to $H^{k,k}$ for $k \neq n-1$).
We will see that the remaining state
arises from the Landau-Ginzburg orbifold (LGO).)

Next, we consider the Landau-Ginzburg ${\mathbb Z}_2$ orbifold.
Here, for later use, note $\rho \in {\mathbb Z}_{2(n-1)}$,
whose generator we label $\zeta$.  The states are as follows:
\begin{itemize}
\item $r=0$:  one state of charge $(q_L,q_R) = (0,0)$,
$\rho$ eigenvalue $1$,
which contributes to $H^{k,\ell}$ with $\ell-k = 0$ and
$k + \ell \equiv 0 \mod 2(n-1)$, hence a linear combination of 
$H^{p,p}$ for $2p \equiv 0 \mod 2(n-1)$.

\item $r=1$:
states of the form
\begin{equation}
{\cal U}_{-1/2,1/2} \, f_p(x) | 0, {\rm RR} \rangle
\end{equation}
with $f_p \sim f_p + dW$ for $p$ satisfying~(\ref{eq:constr-p}).
Since the ideal $(dW)$ is generated by linear monomials $x_i$,
this can only receive contributions from states with $p=0$,
which will only happen if $(n+1)/2 \in {\mathbb Z}$, 
meaning $n+1$ is even.  
In that case, if $n+1$ is even,
we get one state of this form, with charges $(q_L,q_R) = (0,0)$,
and $\rho$ eigenvalue $\zeta^{n-1}$, where $\zeta$ generates
${\mathbb Z}_{2(n-1)}$.  This state contributes to $H^{k,\ell}$
for $\ell-k = 0$, $k+\ell = n-1 \mod 2(n-1)$, hence $H^{(n-1)/2, (n-1)/2}$.
\end{itemize}

Thus, if $n+1$ is even, the Landau-Ginzburg orbifold contributes two
states (one from each of $r=0$ and $r=1$), 
and if $n+1$ is odd, it only contributes one (from $r=0$).
This matches the results in \cite[section 5.3]{Gu:2020zpg},
\cite{Hellerman:2006zs,Caldararu:2007tc,Hori:2011pd},
and also completes the description of the physical origin of the
cohomology.

\subsection{Example: cubic hypersurfaces}

Next, we describe the contributions from the Landau-Ginzburg orbifold
for a cubic hypersurface in ${\mathbb P}^n$.

From equation~(\ref{eq:midcoh:dim}),
\begin{equation}
D_p(n,3) \: = \: \frac{2}{3} \left( 2^n \: + \: (-)^{n-1} \right).
\end{equation}
Expanding the generating function~(\ref{eq:gen-fn-hodge}), we have
\begin{eqnarray}
\sum_{k,\ell} \left( h^{k,\ell} - \delta_{k,\ell} \right)
x^k y^{\ell}
& = &
\frac{
(1+y)^2 - (1+x)^2
}{
(1+x)^3 y - (1+y)^3 x
},
\\
& = & 
\frac{ 2 + x + y }{ 1  - x^2 y - 3 x y - xy^2 },
\\
& = & 
 (2 + x + y) \sum_{k=0}^{\infty} \left( x^2 y + 3 x y + x y^2 \right)^k,
\\
& = &
 2\: + \: (x+y) \: + \: 2 (x^2 y + 3 x y + x y^2 )
\nonumber \\
& & 
\: + \: (x+y) ( x^2 y + 3 x y + x y^2) \: + \: \cdots
\\
& = &
2 \: + \: (x+y) \: + \: (6 x y) \: + \: ( 5 x^2 y + 5 x y^2) \: + \: \cdots.
\end{eqnarray}

We list some special cases below, comparing results from both
physics and mathematics.  In each case, we will take the Landau-Ginzburg
superpotential to be of Fermat type, meaning
\begin{equation}
W \: = \: \sum_{i=0}^n x_i^d.
\end{equation}
Note that in the $r=1$ sector, since for a Fermat cubic the ideal
$(dW)$ includes all degree-two terms of the form $(x_i)^2$, the number
of surviving $f_p$, modulo the ideal $(dW)$, for any $p$ is
\begin{equation}
\left( \begin{array}{c} n+1 \\ p \end{array} \right)
\: = \:
\frac{ (n+1)! }{ (n+1-p)! \, p! }.
\end{equation}

In each case, we will begin by listing mathematics results for the
Dolbeault cohomology groups, and then give the corresponding physics.
We will count states arising from both the Coulomb and the Higgs
(Landau-Ginzburg orbifold) branches.
For the latter, we will 
use symmetries to determine which Dolbeault groups
$H^{k,\ell}$ the states should contribute to, using
the relation~(\ref{eq:dol-1})
\begin{equation}
\ell - k \: = \: q_L + q_R,
\end{equation}
and the fact that the $\rho$ eigenvalue determines
$k + \ell \mod 2(n+1-d)$,
using the $\rho$ action in equation~(\ref{eq:rho-action}).

\subsubsection{$n=3$}

In this section we consider\footnote{
In passing, we note that this example lies outside the bound~(\ref{eq:n-constr}).
} ${\mathbb P}^3[3]$.
Mathematically, $D_p(n,3) = 6$, and
expanding the generating function, for fixed $n=3$, we have
\begin{equation}
\sum_{k,\ell} \left( h^{k,\ell} - \delta_{k,\ell} \right)
x^k y^{\ell}
\: = \: 6 x y,
\end{equation}
from which we deduce
\begin{equation}
h^{1,1} = 7, 
\end{equation}
plus
\begin{equation}
h^{0,0} \: = \: 1 \: = \: h^{2,2}.
\end{equation}

Now, let us turn to the corresponding physics.
From the Coulomb branch, there are $\sigma$ fields, obeying the
relation~(\ref{eq:qc:ci}) 
\begin{equation}
\sigma^{4-1} \: = \: q (-3)^3 \sigma^2,
\end{equation}
or equivalently for our purposes
\begin{equation}
\sigma \: = \: (-3)^3 q.
\end{equation}
The Coulomb branch only contributes one state to
the Dolbeault cohomology groups $H^{p,p}$, which will be a linear
combination of that one state and certain Landau-Ginzburg orbifold states.

Applying our earlier analysis, the Landau-Ginzburg orbifold contributes
\begin{itemize}
\item $r=0$: one state of charge $(q_L,q_R) = (0,0)$
and $\rho$ eigenvalue $1$,
\item $r=1$: six states arising in the RR sector in the form
\begin{equation}
{\cal U}_{-1/2,1/2} \, f_p(x) | 0, {\rm RR} \rangle
\end{equation}
for $p=2$ (satisfying~(\ref{eq:constr-p})) and $f_p \sim f_p + dW$, meaning that the $f_2$ are linear
combinations of
\begin{equation}
x_0 x_1, \: \: \:
x_0 x_2, \: \: \:
x_0 x_3, \: \: \:
x_1 x_2, \: \: \:
x_1 x_3, \: \: \:
x_2 x_3
\end{equation}
of charge $(q_L, q_R) = (-2/3,+2/3)$ and $\rho$ eigenvalue $1$,
\item $r=2$: one state of charge $(q_L,q_R) = (-4/3,+4/3)$ and
$\rho$ eigenvalue $1$.
\end{itemize}

Now, let us interpret these Landau-Ginzburg states mathematically.
First, recall states contribute to $H^{p,q}$ for $q-p = q_L + q_R$,
but all the Landau-Ginzburg states have $q_L + q_R = 0$, hence
they all contribute to $H^{p,p}$ for various $p$.  Also,
they all have $\rho$ eigenvalue $+1$, and as $\rho$ generates a ${\mathbb Z}_2$,
this means they all are in even-degree cohomology -- as is true of all
of the cohomology of ${\mathbb P}^3[3]$.
As a result, in this case we cannot distinguish which Landau-Ginzburg states
contribute to which precise Dolbeault cohomology groups; however, it is clear
that the total collection of Landau-Ginzburg states plus the Coulomb branch
state spans the dimension of the
cohomology groups, as expected.

We summarize our results below on a Hodge diamond, using
$\sigma$ to indicate Coulomb-branch contributions from $\sigma$ fields,
and LG to indicate Higgs-branch contributions from Landau-Ginzburg
orbifold states:
\begin{center}
\begin{tabular}{ccccccccccccccccc}
& & $h^{0,0}$ & &    & &    & & 1 & &     & &      & & LG$_{r=0,1,2}$, $\sigma$ & & \\
& $h^{1,0}$ & & $h^{0,1}$ &  & &    & 0 & & 0 &   & &      & 0 & & 0 & \\
$h^{2,0}$ & & $h^{1,1}$ & & $h^{0,2}$   & $=$  &   
0 & & 7 & & 0  & $=$ &  0 & & LG$_{r=0,1,2}$, $\sigma$ & & 0 \\
& $h^{2,1}$ & & $h^{1,2}$ &   & &     & 0 & & 0 &   & &      & 0 & & 0 & \\
& & $h^{2,2}$ & &     & &     & & 1 & &     & &      & & LG$_{r=0,1,2}$, $\sigma$ & &
\end{tabular}
\end{center}

\subsubsection{$n=4$}

In this section we consider ${\mathbb P}^4[5]$.
Mathematically, $D_p(n,3) = 10$, and
expanding the generating function, for fixed $n=4$, we have
\begin{eqnarray}
\sum_{k,\ell} \left( h^{k,\ell} - \delta_{k,\ell} \right)
x^k y^{\ell}
& =  & 
 5 x^2 y + 5 x y^2,
\end{eqnarray}
from which we deduce
\begin{equation}
h^{2,1} = 5, \: \: \:
h^{1,2} = 5, 
\end{equation}
plus
\begin{equation}
h^{0,0} \: = \: 1 \: = \: h^{1,1} \: = \: h^{2,2} \: = \: h^{3,3}.
\end{equation}

Now, let us turn to the corresponding physics.
From the Coulomb branch, there are $\sigma$ fields, obeying the
relation~(\ref{eq:qc:ci}) 
\begin{equation}
\sigma^{5-1}  \: = \: q (-3)^3 \sigma^2,
\end{equation}
or equivalently for our purposes
\begin{equation}
\sigma^2 \: = \:  (-3)^3 q.
\end{equation}
The Coulomb branch contributes two states to the
Dolbeault cohomology groups $H^{p,p}$.
(We will see that the remainder arise from Landau-Ginzburg orbifold states.)

Next, we consider the Landau-Ginzburg orbifold.
Here, for later use, note $\rho \in {\mathbb Z}_4$,
whose generator we label $\zeta$.  The states are as follows:
\begin{itemize}
\item $r=0$:  one state of charge $(q_L,q_R) = (0,0)$,
$\rho$ eigenvalue $1$,
which contributes to $H^{k,\ell}$ with $\ell-k = 0$ and
$k + \ell \equiv 0 \mod 4$, hence a linear combination of $H^{0,0}$,
$H^{2,2}$.
\item $r=1$: 
states of the form
\begin{equation}
{\cal U}_{-1/2,1/2} \, f_p(x) | 0, {\rm RR} \rangle
\end{equation}
with $f_p \sim f_p + dW$ for $p$ satisfying~(\ref{eq:constr-p}):  
\begin{itemize}
\item five states for $p=1$ of the form $f_p = x_i$ of charge 
$(q_L,q_R) = (-4/3,+1/3)$, $\rho$ eigenvalue $\zeta^3$,
which contribute to $H^{k,\ell}$ with $\ell-k = -1$ 
and $k + \ell \equiv 3 \mod 4$,
hence $H^{2,1}$,
\item five states for $p=4$ of the form $f_p = x_i x_j x_k x_{\ell}$ 
of charge $(q_L,q_R) = (-1/3,+4/3)$, $\rho$ eigenvalue $\zeta^3$,
which contribute to $H^{k,\ell}$ with $\ell-k = +1$
and $k+\ell \equiv 3 \mod 4$,
hence $H^{1,2}$.
\end{itemize}
\item $r=2$: one state of charge $(q_L,q_R) = (-5/3,+5/3)$,
$\rho$ eigenvalue $\zeta^6$,
which contribute to $H^{k,\ell}$ with $\ell-k=0$ and $k+\ell \equiv 2 \mod 4$,
hence a linear combination of $H^{1,1}$, $H^{3,3}$.
\end{itemize}

We summarize our results below on a Hodge diamond, using
$\sigma$ to indicate Coulomb-branch contributions from $\sigma$ fields,
and LG to indicate Higgs-branch contributions from Landau-Ginzburg
orbifold states:
\begin{center}
\begin{tabular}{ccccccccccccccc}
& & & 1 & & &   & &  & & & LG$_{r=0}$, $\sigma$ & & & \\
& & 0 & & 0 & &  & &   & & 0 & & 0 & & \\
& 0 & & 1 & & 0 &   & &    & 0 & & LG$_{r=2}$, $\sigma$ & & 0 &  \\
0 & & 5 & & 5 & & 0 &  $=$ &  0 & & LG & & LG & & 0 \\
& 0 & & 1 & & 0 &   & &    & 0 & & LG$_{r=0}$, $\sigma$ & & 0 &  \\
& & 0 & & 0 & &  & &   & & 0 & & 0 & & \\
& & & 1 & & &   & &  & & & LG$_{r=2}$, $\sigma$ & & &
\end{tabular}
\end{center}

\subsubsection{$n=5$}

In this section we consider ${\mathbb P}^5[3]$.
Mathematically, $D_p(n,3) = 22$, and
expanding the generating function, for fixed $n=5$, we have
\begin{equation}
\sum_{k,\ell} \left( h^{k,\ell} - \delta_{k,\ell} \right)
x^k y^{\ell}
\: = \:
x^3 y + 20 x^2 y^2 + x y^3,
\end{equation}
from which we deduce
\begin{equation}
h^{3,1} = 1, \: \: \:
h^{2,2} = 21, \: \: \:
h^{1,3} = 1,
\end{equation}
plus
\begin{equation}
h^{0,0} \: = \: 1 \: = \: h^{1,1} \: = \: h^{2,2} \: = \: h^{3,3} \: = \: h^{4,4}.
\end{equation}

Now, let us turn to the corresponding physics.
From the Coulomb branch, there are $\sigma$ fields, obeying the
relation~(\ref{eq:qc:ci})
\begin{equation}
\sigma^{6-1}  \: = \: q (-3)^3 \sigma^2,
\end{equation}
or equivalently for our purposes
\begin{equation}
\sigma^3 \: = \:  (-3)^3 q.
\end{equation}
The Coulomb branch contributes three states to $H^{p,p}$,
and we will see the remainder arise from Landau-Ginzburg orbifold states.

Next we turn to the Landau-Ginzburg orbifold.  In passing, this orbifold
(a ${\mathbb Z}_3$ orbifold of a superpotential that is degree 3 in
six variables, is related to the SCFT for a K3 surface,
implemented as an orbifold of a product of two elliptic curves
${\mathbb P}^2[3]$, consistent with the fact that this theory
flows to a SCFT with $c/3 = 2$, from~(\ref{eq:c}).

Next, we consider the Landau-Ginzburg orbifold states.
Here, for later use, note that $\rho \in {\mathbb Z}_6$,
whose generator we label $\zeta$.  The states are as follows:
\begin{itemize}
\item $r=0$:  one state of charge $(q_L,q_R) = (0,0)$,
$\rho$ eigenvalue $1$,
which contributes to $H^{k,\ell}$ with $\ell-k = 0$ and
$k+\ell \equiv 0 \mod 6$, hence
a linear combination of $H^{0,0}$, $H^{3,3}$.
\item $r=1$: 
states of the form
\begin{equation}
{\cal U}_{-1/2,1/2} \, f_p(x) | 0, {\rm RR} \rangle
\end{equation}
with $f_p \sim f_p + dW$ for $p$ satisfying~(\ref{eq:constr-p}): 
\begin{itemize}
\item one state for $p=0$ of the form $f_p = 1$ of charge 
$(q_L,q_R) = (-2,0)$, $\rho$ eigenvalue $\zeta^4$,
which contribute to $H^{k,\ell}$ with $\ell-k = -2$
and $k+\ell \equiv 4 \mod 6$,
hence $H^{3,1}$.
\item 20 states for $p=3$ of the form $f_p = x_i x_j x_k$ of charge 
$(q_L,q_R) = (-1,+1)$, $\rho$ eigenvalue $\zeta^4$,
which contribute to $H^{k,\ell}$ with $\ell-k = 0$
and $k+\ell \equiv 4 \mod 6$,
hence $H^{2,2}$.
\item one state for $p=6$ of the form 
$f_p = x_{i_1} x_{i_2} x_{i_3} x_{i_4} x_{i_5} x_{i_6}$ of charge 
$(q_L,q_R) = (0,+2)$, $\rho$ eigenvalue $\zeta^4$,
which contributes to $H^{k,\ell}$ with $\ell-k = +2$ and
$k+\ell \equiv 4 \mod 6$,
hence $H^{1,3}$.
\end{itemize}
\item $r=2$: one state of charge $(q_L,q_R) = (-2,+2)$,
$\rho$ eigenvalue $\zeta^8$,
which contributes to $H^{k,\ell}$ with $\ell-k = 0$ and
$k + \ell \equiv 8 \mod 6$,
hence a linear combination of $H^{1,1}$, $H^{4,4}$.
\end{itemize}

We summarize our results below on a Hodge diamond, using
$\sigma$ to indicate Coulomb-branch contributions from $\sigma$ fields,
and LG to indicate Higgs-branch contributions from Landau-Ginzburg
orbifold states:
\begin{center}
\begin{tabular}{ccccccccccccccccccc}
& & & & 1 & & & &       & &      & & & & LG$_{r=0}$, $\sigma$ & & & &  \\
& & & 0 & & 0 & & &     & &      & & & 0 & & 0 & & &  \\
& & 0 & & 1 & & 0 & &   & &      & & 0 & & LG$_{r=2}$, $\sigma$ & & 0 & &  \\
& 0 & & 0 & & 0 & & 0 & & &      & 0 & & 0 & & 0 & & 0 & \\
0 & & 1 & & 21 & & 1 & & 0   & $=$ &   0 & & LG & & LG$_{r=1}$, $\sigma$ & & LG & & 0 \\
& 0 & & 0 & & 0 & & 0 & & &      & 0 & & 0 & & 0 & & 0 & \\
& & 0 & & 1 & & 0 & &   & &      & & 0 & & LG$_{r=0}$, $\sigma$ & & 0 & &  \\
& & & 0 & & 0 & & &     & &      & & & 0 & & 0 & & &  \\
& & & & 1 & & & &       & &      & & & & LG$_{r=2}$, $\sigma$ & & & &
\end{tabular}
\end{center}

\subsubsection{$n=6$}

In this section we consider ${\mathbb P}^6[3]$.
Mathematically, $D_p(n,3) = 42$, and
expanding the generating function, for fixed $n=6$, we have
\begin{equation}
\sum_{k,\ell} \left( h^{k,\ell} - \delta_{k,\ell} \right)
x^k y^{\ell}
\: = \:
21 x^3 y^2 + 21 x^2 y^3,
\end{equation}
from which we deduce
\begin{equation}
h^{3,2} = 21, \: \: \: h^{2,3} = 21,
\end{equation}
plus
\begin{equation}
h^{0,0} \: = \: 1 \: = \: h^{1,1} \: = \: h^{2,2} \: = \: h^{3,3}
\: = \: h^{4,4} \: = \: h^{5.5}.
\end{equation}

Now, let us turn to the corresponding physics.
From the Coulomb branch, there are $\sigma$ fields, obeying the
relation~(\ref{eq:qc:ci})
\begin{equation}
\sigma^{7-1}  \: = \: q (-3)^3 \sigma^2,
\end{equation}
or equivalently for our purposes
\begin{equation}
\sigma^4 \: = \:  (-3)^3 q.
\end{equation}
The Coulomb branch contributes four states to $H^{k,k}$,
and we will see the remainder arise from Landau-Ginzburg orbifold states.

Next we turn to the Landau-Ginzburg orbifold. 
Here, for later use, note that $\rho \in {\mathbb Z}_8$,
whose generator we label $\zeta$.  The states are as follows:
\begin{itemize}
\item $r=0$:  one state of charge $(q_L,q_R) = (0,0)$,
$\rho$ eigenvalue $1$, 
which contributes to $H^{k,\ell}$ for $\ell-k = 0$
and $k+\ell \equiv 0 \mod 8$, hence
a linear combination of $H^{0,0}$, $H^{4,4}$.
\item $r=1$:
states of the form
\begin{equation}
{\cal U}_{-1/2,1/2} \, f_p(x) | 0, {\rm RR} \rangle
\end{equation}
with $f_p \sim f_p + dW$ for $p$ satisfying~(\ref{eq:constr-p}): 
\begin{itemize}
\item 21 states for $p=2$ of the form $f_p = x_i x_j$ of charge 
$(q_L,q_R) = (-5/3, +2/3)$, $\rho$ eigenvalue $\zeta^5$,
which contribute to $H^{k,\ell}$ for $\ell-k = -1$ and
$k+\ell \equiv 5 \mod 8$, hence
$H^{3,2}$.
\item 21 states for $p=5$ of the form $f_p = x_{i_1} x_{i_2} x_{i_3} x_{i_4} x_{i_5}$ 
of charge $(q_L,q_R) = (-2/3, 5/3)$, $\rho$ eigenvalue $\zeta^5$,
which contribute to $H^{k,\ell}$ for $\ell-k = +1$ and
$k+\ell \equiv 5 \mod 8$,
hence $H^{2,3}$.
\end{itemize}
\item $r=2$: one state of charge $(q_L,q_R) = (-7/3, +7/3)$,
$\rho$ eigenvalue $\zeta^{10}$,
which contributes to $H^{k,\ell}$ for $\ell-k = 0$ and
$k+\ell \equiv 10 \mod 8$,
hence a linear combination of $H^{1,1}$, $H^{5,5}$.
\end{itemize}

In terms of the Hodge diamond for ${\mathbb P}^6[3]$,
\begin{center}
\begin{tabular}{ccccccccccc}
& & & & & 1 & & & & & \\
& & & & 0 & & 0 & & & & \\
& & & 0 & & 1 & & 0 & & & \\
& & 0 & & 0 & & 0 & & 0 & & \\
& 0 & & 0 & & 1 & & 0 & & 0 & \\
0 & & 0 & & 21 & & 21 & & 0 & & 0 \\
& 0 & & 0 & & 1 & & 0 & & 0 & \\
& & 0 & & 0 & & 0 & & 0 & & \\
& & & 0 & & 1 & & 0 & & & \\
& & & & 0 & & 0 & & & & \\
& & & & & 1 & & & & &
\end{tabular}
\end{center}
the physical origin of the cohomology is described as
\begin{center}
\begin{tabular}{ccccccccccc}
& & & & & LG$_{r=0}$, $\sigma$ & & & & & \\
& & & & 0 & & 0 & & & & \\
& & & 0 & & LG$_{r=2}$, $\sigma$ & & 0 & & & \\
& & 0 & & 0 & & 0 & & 0 & & \\
& 0 & & 0 & & $\sigma$ & & 0 & & 0 & \\
0 & & 0 & & LG & & LG & & 0 & & 0 \\
& 0 & & 0 & & $\sigma$ & & 0 & & 0 & \\
& & 0 & & 0 & & 0 & & 0 & & \\
& & & 0 & & LG$_{r=0}$, $\sigma$ & & 0 & & & \\
& & & & 0 & & 0 & & & & \\
& & & & & LG$_{r=2}$, $\sigma$ & & & & &
\end{tabular}
\end{center}

\subsubsection{$n=7$}

In this section we consider ${\mathbb P}^7[3]$.
Mathematically, $D_p(n,3) = 86$, and
expanding the generating function, for fixed $n=7$, we have
\begin{equation}
\sum_{k,\ell} \left( h^{k,\ell} - \delta_{k,\ell} \right)
x^k y^{\ell}
\: = \:
8 x^4 y^2 + 70 x^3 y^3 + 8 x^2 y^4,
\end{equation}
from which we deduce
\begin{equation}
h^{4,2} = 8, \: \: \: h^{3,3} = 71, \: \: \: h^{2,4} = 8,
\end{equation}
plus
\begin{equation}
h^{0,0} \: = \: 1 \: = \: h^{1,1} \: = \: h^{2,2} 
\: = \: h^{4,4} \: = \: h^{5.5} \: = \: h^{6,6}.
\end{equation}

Now, let us turn to the corresponding physics.
From the Coulomb branch, there are $\sigma$ fields, obeying the
relation~(\ref{eq:qc:ci})
\begin{equation}
\sigma^{8-1}  \: = \: q (-3)^3 \sigma^2,
\end{equation}
or equivalently for our purposes
\begin{equation}
\sigma^5 \: = \:  (-3)^3 q.
\end{equation}
The Coulomb branch contributes five states to $H^{k,k}$,
and we will see that the remainder arise from Landau-Ginzburg orbifold
states.

Next we turn to the Landau-Ginzburg orbifold.
Here, for later use, note that $\rho \in {\mathbb Z}_{10}$,
whose generator we label $\zeta$.  The states are as follows:
\begin{itemize}
\item $r=0$:  one state of charge $(q_L,q_R) = (0,0)$,
$\rho$ eigenvalue $1$,
which contributes to $H^{k,\ell}$ for $\ell-k = 0$ and
$k+\ell \equiv 0 \mod 10$,
hence a linear combination of $H^{0,0}$, $H^{5,5}$.
\item $r=1$:
states of the form
\begin{equation}
{\cal U}_{-1/2,1/2} \, f_p(x) | 0, {\rm RR} \rangle
\end{equation}
with $f_p \sim f_p + dW$ for $p$ satisfying~(\ref{eq:constr-p}):
\begin{itemize}
\item 8 states for $p=1$ of the form $f_p = x_i$ of charge 
$(q_L,q_R) = (-7/3, +1/3)$, $\rho$ eigenvalue $\zeta^6$,
which contribute to $H^{k,\ell}$ for $\ell-k = -2$
and $k+\ell \equiv 6 \mod 10$, hence $H^{4,2}$.
\item 70 states for $p=4$ of the form $f_p = x_{i_1} x_{i_2} x_{i_3} x_{i_4}$ 
of charge $(q_L,q_R) = (-4/3, +4/3)$,
$\rho$ eigenvalue $\zeta^6$,
which contribute to $H^{k,\ell}$ for $\ell-k = 0$ and
$k+\ell \equiv 6 \mod 10$,
hence $H^{3,3}$.
\item 8 states for $p=7$ of the form 
$f_p = x_{i_1} x_{i_2} x_{i_3} x_{i_4} x_{i_5} x_{i_6} x_{i_7}$ 
of charge $(q_L,q_R) = (-1/3, +7/3)$,
$\rho$ eigenvalue $\zeta^6$
which contribute to $H^{k,\ell}$ for $\ell-k = +2$ and
$k+\ell \equiv 6 \mod 10$,
hence $H^{2,4}$.
\end{itemize}
\item $r=2$: one state of charge $(q_L,q_R) = (-8/3, +8/3)$,
$\rho$ eigenvalue $\zeta^{12}$,
which contributes to $H^{k,\ell}$ for $\ell-k=0$ and
$k+\ell \equiv 12 \mod 10$,
hence a linear combination of $H^{1,1}$, $H^{6,6}$.
\end{itemize}

In terms of the Hodge diamond for ${\mathbb P}^7[3]$,
\begin{center}
\begin{tabular}{ccccccccccccc}
& & & & & & 1 & & & & & & \\
& & & & & 0 & & 0 & & & & & \\
& & & & 0 & & 1 & & 0 & & & & \\
& & & 0 & & 0 & & 0 & & 0 & & & \\
& & 0 & & 0 & & 1 & & 0 & & 0 & & \\
& 0 & & 0 & & 0 & & 0 & & 0 & & 0 & \\
0 & & 0 & & 8 & & 71 & & 8 & & 0 & & 0 \\
& 0 & & 0 & & 0 & & 0 & & 0 & & 0 & \\
& & 0 & & 0 & & 1 & & 0 & & 0 & & \\
& & & 0 & & 0 & & 0 & & 0 & & & \\
& & & & 0 & & 1 & & 0 & & & & \\
& & & & & 0 & & 0 & & & & & \\
& & & & & & 1 & & & & & &
\end{tabular}
\end{center}
the physical origin of the cohomology is described as
\begin{center}
\begin{tabular}{ccccccccccccc}
& & & & & & LG$_{r=0}$, $\sigma$ & & & & & & \\
& & & & & 0 & & 0 & & & & & \\
& & & & 0 & & LG$_{r=2}$, $\sigma$ & & 0 & & & & \\
& & & 0 & & 0 & & 0 & & 0 & & & \\
& & 0 & & 0 & & $\sigma$ & & 0 & & 0 & & \\
& 0 & & 0 & & 0 & & 0 & & 0 & & 0 & \\
0 & & 0 & & LG & & LG$_{r=1}$, $\sigma$ & & LG & & 0 & & 0 \\
& 0 & & 0 & & 0 & & 0 & & 0 & & 0 & \\
& & 0 & & 0 & & $\sigma$ & & 0 & & 0 & & \\
& & & 0 & & 0 & & 0 & & 0 & & & \\
& & & & 0 & & LG$_{r=0}$, $\sigma$ & & 0 & & & & \\
& & & & & 0 & & 0 & & & & & \\
& & & & & & LG$_{r=2}$, $\sigma$ & & & & & &
\end{tabular}
\end{center}

\subsubsection{$n=8$}

In this section we consider ${\mathbb P}^8[3]$.
Mathematically, $D_p(n,3) = 170$, and
expanding the generating function, for fixed $n=8$, we have
\begin{equation}
\sum_{k,\ell} \left( h^{k,\ell} - \delta_{k,\ell} \right)
x^k y^{\ell}
\: = \:
x^5 y^2 + 84 x^4 y^3 + 84 x^3 y^4 + x^2 y^5,
\end{equation}
from which we deduce
\begin{equation}
h^{5,2} = 1, \: \: \:
h^{4,3} = 84, \: \: \:
h^{3,4} = 84, \: \: \:
h^{2,5} = 1,
\end{equation}
plus
\begin{equation}
h^{0,0} \: = \: 1 \: = \: h^{1,1} \: = \: h^{2,2} \: = \: h^{3,3}
\: = \: h^{4,4} \: = \: h^{5.5} \: = \: h^{6,6} \: = \: h^{7,7}.
\end{equation}

Now, let us turn to the corresponding physics.
From the Coulomb branch, there are $\sigma$ fields, obeying the
relation~(\ref{eq:qc:ci})
\begin{equation}
\sigma^{9-1}  \: = \: q (-3)^3 \sigma^2,
\end{equation}
or equivalently for our purposes
\begin{equation}
\sigma^6 \: = \:  (-3)^3 q.
\end{equation}
The Coulomb branch contributes six states to
$H^{k,k}$, and we will see that the remainder are contributed
by Landau-Ginzburg orbifold states.

Next we turn to the Landau-Ginzburg orbifold.
Here, for later use, note that $\rho \in {\mathbb Z}_{12}$,
whose generator we label $\zeta$.  The states are as follows:
\begin{itemize}
\item $r=0$:  one state of charge $(q_L,q_R) = (0,0)$,
$\rho$ eigenvalue $1$,
which contributes to $H^{k,\ell}$ for $\ell-k = 0$
and $k+\ell \equiv 0 \mod 12$,
hence a linear combination of $H^{0,0}$, $H^{6,6}$.
\item $r=1$:
states of the form
\begin{equation}
{\cal U}_{-1/2,1/2} \, f_p(x) | 0, {\rm RR} \rangle
\end{equation}
with $f_p \sim f_p + dW$ for $p$ satisfying~(\ref{eq:constr-p}):
\begin{itemize}
\item one state for $p=0$ of the form $f_p = 1$ of charge 
$(q_L,q_R) = (-3,0)$, $\rho$ eigenvalue $\zeta^7$,
which contributes to $H^{k,\ell}$ for $\ell-k = -3$ and
$k+\ell \equiv 7 \mod 12$,
hence $H^{5,2}$.
\item 84 states for $p=3$ of the form $f_p = x_{i_1} x_{i_2} x_{i_3}$ 
of charge $(q_L,q_R) = (-2,+1)$, $\rho$ eigenvalue $\zeta^7$,
which contribute to $H^{k,\ell}$ for $\ell-k = -1$ and
$k+\ell \equiv 7 \mod 12$,
hence $H^{4,3}$.
\item 84 states for $p=6$ of the form 
$f_p = x_{i_1} x_{i_2} x_{i_3} x_{i_4} x_{i_5} x_{i_6}$ 
of charge $(q_L,q_R) = (-1,+2)$, $\rho$ eigenvalue $\zeta^7$,
which contribute to $H^{k,\ell}$ for $\ell-k = +1$ and
$k+\ell \equiv 7 \mod 12$,
hence $H^{3,4}$.
\item one state for $p=9$ of the form 
$f_p = x_{i_1} x_{i_2} x_{i_3} x_{i_4} x_{i_5} x_{i_6} x_{i_7} x_{i_8} x_{i_9}$ 
of charge $(q_L,q_R) = (0,+3)$, $\rho$ eigenvalue $\zeta^7$,
which contributes to $H^{k,\ell}$ for $\ell-k = +3$ and
$k+\ell \equiv 7 \mod 12$,
hence $H^{2,5}$.
\end{itemize}
\item $r=2$: one state of charge $(q_L,q_R) = (-3,+3)$,
$\rho$ eigenvalue $\zeta^{14}$,
which contributes to $H^{k,\ell}$ for $\ell-k=0$ and
$k+\ell \equiv 14 \mod 12$,
hence a linear combination of $H^{1,1}$, $H^{7,7}$.
\end{itemize}

In terms of the Hodge diamond for ${\mathbb P}^8[3]$,
\begin{center}
\begin{tabular}{ccccccccccccccc}
& & & & & & & 1 & & & & & & & \\
& & & & & & 0 & & 0 & & & & & & \\
& & & & & 0 & & 1 & & 0 & & & & & \\
& & & & 0 & & 0 & & 0 & & 0 & & & & \\
& & & 0 & & 0 & & 1 & & 0 & & 0 & & & \\
& & 0 & & 0 & & 0 & & 0 & & 0 & & 0 & & \\
& 0 & & 0 & & 0 & & 1 & & 0 & & 0 & & 0 & \\
0 & & 0 & & 1 & & 84 & & 84 & & 1 & & 0 & & 0 \\
& 0 & & 0 & & 0 & & 1 & & 0 & & 0 & & 0 & \\
& & 0 & & 0 & & 0 & & 0 & & 0 & & 0 & & \\
& & & 0 & & 0 & & 1 & & 0 & & 0 & & & \\
& & & & 0 & & 0 & & 0 & & 0 & & & & \\
& & & & & 0 & & 1 & & 0 & & & & & \\
& & & & & & 0 & & 0 & & & & & & \\
& & & & & & & 1 & & & & & & &
\end{tabular}
\end{center}
the physical origin of the cohomology is described as
\begin{center}
\begin{tabular}{ccccccccccccccc}
& & & & & & & LG$_{r=0}$, $\sigma$ & & & & & & & \\
& & & & & & 0 & & 0 & & & & & & \\
& & & & & 0 & & LG$_{r=2}$, $\sigma$ & & 0 & & & & & \\
& & & & 0 & & 0 & & 0 & & 0 & & & & \\
& & & 0 & & 0 & & $\sigma$ & & 0 & & 0 & & & \\
& & 0 & & 0 & & 0 & & 0 & & 0 & & 0 & & \\
& 0 & & 0 & & 0 & & $\sigma$ & & 0 & & 0 & & 0 & \\
0 & & 0 & & LG & & LG & & LG & & LG & & 0 & & 0 \\
& 0 & & 0 & & 0 & & $\sigma$ & & 0 & & 0 & & 0 & \\
& & 0 & & 0 & & 0 & & 0 & & 0 & & 0 & & \\
& & & 0 & & 0 & & $\sigma$ & & 0 & & 0 & & & \\
& & & & 0 & & 0 & & 0 & & 0 & & & & \\
& & & & & 0 & & LG$_{r=0}$, $\sigma$ & & 0 & & & & & \\
& & & & & & 0 & & 0 & & & & & & \\
& & & & & & & LG$_{r=2}$, $\sigma$ & & & & & & &
\end{tabular}
\end{center}

\subsection{Example:  degree four hypersurfaces}

From equation~(\ref{eq:midcoh:dim}),
\begin{equation}
D_p(n,3) \: = \: \frac{3}{4}\left( 3^n + (-)^{n-1} \right).
\end{equation}
Expanding the generating function~(\ref{eq:gen-fn-hodge}), we have
\begin{eqnarray}
\sum_{k,\ell} \left( h^{k,\ell} - \delta_{k,\ell} \right)
x^k y^{\ell}
& = &
\frac{
(1+y)^3 - (1+x)^3
}{
(1+x)^4 y - (1+y)^4 x
},
\\
& = &
\frac{
3 + 3 (y+x) + (x^2 + xy + y^2)
}{
1 - 6 xy - 4 xy(y+x) - xy(x^2 + xy + y^2)
}.
\end{eqnarray}

We list some special cases below, comparing results from both physics
and mathematics.  As in the degree three examples, we will take the
Landau-Ginzburg superpotential to be of Fermat type, meaning
\begin{equation}
W \: = \: \sum_{i=0}^{n} x_i^d.
\end{equation}
In each case, we will begin by listing mathematics results for the
Dolbeault cohomology groups, and then give the corresponding physics.
As before, we will count states arising from both the Coulomb and
the Higgs (Landau-Ginzburg orbifold) branches, and in the latter, 
determine which Dolbeault groups $H^{k,\ell}$ the
states should contribute to 
using
the relation~(\ref{eq:dol-1})
\begin{equation}
\ell - k \: = \: q_L + q_R,
\end{equation}
and the fact that the $\rho$ eigenvalue determines
$k + \ell \mod 2(n+1-d)$,
using the $\rho$ action in equation~(\ref{eq:rho-action}).

\subsubsection{$n=4$}

In this section we consider\footnote{
In passing, we observe that this example lies outside the bound~(\ref{eq:n-constr}).
} ${\mathbb P}^4[4]$.
Mathematically, $D_p(n,4) = 60$, and
expanding the generating function, for fixed $n=4$, we have
\begin{equation}
\sum_{k,\ell} \left( h^{k,\ell} - \delta_{k,\ell} \right)
x^k y^{\ell}
\: = \:
30 x^2 y + 30 x y^2,
\end{equation}
from which we deduce
\begin{equation}
h^{2,1} = 30, \: \: \: h^{1,2} = 30.
\end{equation}

Now, let us turn to the corresponding physics.
From the Coulomb branch, there are $\sigma$ fields, obeying the
relation~(\ref{eq:qc:ci})
\begin{equation}
\sigma^{5-1} \: = \: q (-4)^4 \sigma^3,
\end{equation}
or equivalently for our purposes
\begin{equation}
\sigma \: = \: (-4)^4 q.
\end{equation}
The Coulomb branch only contributes one state to $H^{k,k}$.
We will see that the remainder are contributed by the Landau-Ginzburg
orbifold.

Next, we consider the Landau-Ginzburg orbifold.
Here, for later use, note $\rho \in {\mathbb Z}_2$,
whose generator we label $\zeta$.  The states are as follows:
\begin{itemize}
\item $r=0$:  one state of charge $(q_L,q_R) = (0,0)$,
$\rho$ eigenvalue $1$, 
which contributes to $H^{k,\ell}$ with $\ell-k=0$ and
$k+\ell \equiv 0 \mod 2$, 
hence $H^{k,k}$.
\item $r=1$:
states of the form
\begin{equation}
{\cal U}_{-1/2,1/2} \, f_p(x) | 0, {\rm RR} \rangle
\end{equation}
with $f_p \sim f_p + dW$, for $p= 3, 7$, satisfying~(\ref{eq:constr-p}).
\begin{itemize}
\item For $p=3$, possible $f_p$ (modulo the ideal) are of the form
\begin{equation}
x_i x_j x_k, \: \: \: x_i x_j^2,
\end{equation}
for $i \neq j \neq k$, of which there are 30 possible terms.
The corresponding states have charge $(q_L,q_R) = (-7/4, +3/4)$
and $\rho$ eigenvalue $\zeta^3$,
so they contribute to $H^{k,\ell}$ for $\ell-k = -1$ and
$k+\ell \equiv 1 \mod 2$,
hence $H^{2,1}$.
\item For $p=7$, possible $f_p$ (modulo the ideal) are of the form
\begin{equation}
x_i^2 x_j^2 x_k^2 x_{\ell}, \: \: \: x_i x_j x_k x_{\ell}^2 x_m^2,
\end{equation}
for $i \neq j \neq k \neq \ell \neq m$, of which there are 30 possible terms.
The corresponding states have charge $(q_L,q_R) = (-3/4, +7/4)$
and $\rho$ eigenvalue $\zeta^3$,
so they contribute to $H^{k,\ell}$ for $\ell-k =+1$ and
$k+\ell \equiv 1 \mod 2$,
hence $H^{1,2}$.
\end{itemize}
\item $r=2$: one state of charge $(q_L,q_R) = (-5/2, +5/2)$,
$\rho$ eigenvalue $1$,
so they contribute to $H^{k,\ell}$ for $\ell-k=0$ and 
$k+\ell \equiv 0 \mod 2$,
hence $H^{k,k}$.
\item $r=3$: one state of charge $(q_L,q_R) = (-5/4, +5/4)$,
$\rho$ eigenvalue $1$,
so they contribute to $H^{k,\ell}$ for $\ell-k=0$ and
$k+\ell \equiv 0 \mod 2$,
hence $H^{k,k}$.
\end{itemize}

The $r=0$, $r=2$, and $r=3$ states could each contribute to any
linear combination of $H^{1,1}$, $H^{2,2}$, $H^{3,3}$; symmetries alone
do not suffice to further distinguish.

We summarize our results below on a Hodge diamond,
\begin{center}
\begin{tabular}{ccccccc}
& & & $h^{0,0}$ & & &   \\
& & $h^{1,0}$ & & $h^{0,1}$ & &  \\
& $h^{2,0}$ & & $h^{1,1}$ & & $h^{0,2}$ &     \\
$h^{3,0}$ & & $h^{2,1}$ & & $h^{1,2}$ & & $h^{0,3}$  \\
& $h^{3,1}$ & & $h^{2,2}$ & & $h^{1,3}$ & \\
& & $h^{3,2}$ & & $h^{2,3}$ & & \\
& & & $h^{3,3}$ & & &
\end{tabular}
\end{center}
using $\sigma$ to indicate Coulomb-branch contributions from $\sigma$
fields, and LG to indicate Higgs-branch contributions from
Landau-Ginzburg-orbifold states:
\begin{center}
\begin{tabular}{ccccccccccccccc}
& & & 1 & & &    & &    & & & LG$_{r=0,2,3}$, $\sigma$ & & &
\\
& & 0 & & 0 & &    & &
& & 0 & & 9 & & 
\\
& 0 & & 1 & & 0 &      & &
& 0 & & LG$_{r=0,2,3}$, $\sigma$ & & 0 &
\\
0 & & 30 & & 30 & & 0       & $=$ & 
0 & & LG & & LG & & 0 \\
& 0 & & 1 & & 0 &      & &
& 0 & & LG$_{r=0,2,3}$, $\sigma$ & & 0 &
\\
& & 0 & & 0 & &    & &
& & 0 & & 0 & & 
\\
& & & 1 & & &    & &    & & & LG$_{r=0,2,3}$, $\sigma$ & & &
\end{tabular}
\end{center}

\subsubsection{$n=5$}

In this section we consider\footnote{
In passing, we observe that this example lies outside the bound~(\ref{eq:n-constr}).
} ${\mathbb P}^5[3]$.
Mathematically, $D_p(n,4) = 183$, and
expanding the generating function, for fixed $n=5$, we have
\begin{equation}
\sum_{k,\ell} \left( h^{k,\ell} - \delta_{k,\ell} \right)
x^k y^{\ell}
\: = \:
21 x^3 y + 141 x^2 y^2 + 21 xy^3,
\end{equation}
from which we deduce
\begin{equation}
h^{3,1} = 21, \: \: \:
h^{2,2} = 142, \: \: \:
h^{1,3} = 21,
\end{equation}
plus
\begin{equation}
h^{0,0} \: = \: 1 \: = \: h^{1,1}
\: = \: h^{3,3} \: = \: h^{4,4}.
\end{equation}

Now, let us turn to the corresponding physics.
From the Coulomb branch, there are $\sigma$ fields, obeying the
relation~(\ref{eq:qc:ci})
\begin{equation}
\sigma^{6-1} \: = \: q (-4)^4 \sigma^3,
\end{equation}
or equivalently for our purposes
\begin{equation}
\sigma^2 \: = \: (-4)^4 q.
\end{equation}
The Coulomb branch only contributes two states to $H^{k,k}$.
We shall see the remainder arise as Landau-Ginzburg orbifold states.

Next, we consider the Landau-Ginzburg orbifold.
Here, for later use, note $\rho \in {\mathbb Z}_4$,
whose generator we label $\zeta$.  The states are as follows:
\begin{itemize}
\item $r=0$:  one state of charge $(q_L,q_R) = (0,0)$,
$\rho$ eigenvalue $1$, which contributes to
$H^{k,\ell}$ with $\ell-k = 0$ and $k+\ell \equiv 0 \mod 4$,
hence to a linear combination of $H^{0,0}$, $H^{2,2}$, $H^{4,4}$.
\item $r=1$:
states of the form
\begin{equation}
{\cal U}_{-1/2,1/2} \, f_p(x) | 0, {\rm RR} \rangle
\end{equation}
with $f_p \sim f_p + dW$, for $p = 2, 6, 10$, satisfying~(\ref{eq:constr-p}).
\begin{itemize}
\item For $p=2$, possible $f_p$ (modulo the ideal) are of the form
\begin{equation}
x_i x_j, \: \: \: x_i^2,
\end{equation}
for $i \neq j$, of which there are 21 possible terms.
The corresponding states have charge $(q_L,q_R) = (-5/2, +1/2)$,
$\rho$ eigenvalue $1$,
and contribute to $H^{k,\ell}$ for $\ell-k = -2$ and
$k+\ell \equiv 0 \mod 4$, hence $H^{3,1}$.
\item For $p=6$, possible $f_p$ (modulo the ideal) are of the form
\begin{equation}
x_i^2 x_j^2 x_k^2, \: \: \:
x_i x_j x_k^2 x_{\ell}^2, \: \: \:
x_i x_j x_k x_{\ell} x_m^2, \: \: \:
x_i x_j x_k x_{\ell} x_m x_n,
\end{equation}
for distinct factors, of which there are 141 possible terms.
The corresponding states have charge $(q_L,q_R) = (-3/2, +3/2)$,
$\rho$ eigenvalue $1$,
and contribute to $H^{k,\ell}$ with $\ell-k = 0$, $k+\ell \equiv 0 \mod 4$,
hence a linear combination of $H^{0,0}$, $H^{2,2}$, and $H^{4,4}$.
\item For $p=10$, possible $f_p$ (modulo the ideal) are of the form
\begin{equation}
x_i^2 x_j^2 x_k^2 x_{\ell}^2 x_m^2, \: \: \:
x_i x_j x_k^2 x_{\ell}^2 x_m^2 x_n^2,
\end{equation}
of which there are 21 possible terms.
The corresponding states have charge $(q_L,q_R) = (-1/2, +5/2)$,
$\rho$ eigenvalue $1$,
and contribute to $H^{k,\ell}$ for $\ell-k = +2$, $k+\ell \equiv 0 \mod 4$,
hence $H^{1,3}$.
\end{itemize}
\item $r=2$: one state of charge $(q_L,q_R) = (-3,+3)$, $\rho$ eigenvalue $1$,
which contributes to $H^{k,\ell}$ with $\ell-k = 0$, $k+\ell \equiv 0 \mod 4$,
hence a linear combination of $H^{0,0}$, $H^{2,2}$, and $H^{4,4}$.
\item $r=3$: one state of charge $(q_L,q_R) = (-3/2, +3/2)$,
$\rho$ eigenvalue $\zeta^2$,
which contributes to $H^{k,\ell}$ with $\ell-k = 0$, $k+\ell \equiv 2 \mod 4$,
hence a linear combination of $H^{1,1}$ and $H^{3,3}$.
\end{itemize}

We summarize our results below on a Hodge diamond,
using $\sigma$ to indicate Coulomb-branch contributions from $\sigma$
fields, and LG to indicate Higgs-branch contributions from
Landau-Ginzburg-orbifold states:
\begin{center}
\begin{tabular}{ccccccccccccccccccc}
& & & & 1 & & & &     & & 
& & & & LG$_{r=0,1,2}$, $\sigma$ & & & & 
\\
& & & 0 & & 0 & & &    & &
& & & 0 & & 0 & & & 
\\
& & 0 & & 1 & & 0 & &    & &
& & 0 & & LG$_{r=3}$, $\sigma$ & & 0 & & 
\\
& 0 & & 0 & & 0 & & 0 &    & &
& 0 & & 0 & & 0 & & 0 &  
\\
0 & & 21 & & 142 & & 21 & & 0     & $=$ &
0 & & LG & & LG$_{r=0,1,2}$, $\sigma$ & & LG & & 0
\\
& 0 & & 0 & & 0 & & 0 &    & &
& 0 & & 0 & & 0 & & 0 &  
\\
& & 0 & & 1 & & 0 & &    & &
& & 0 & & LG$_{r=3}$, $\sigma$ & & 0 & & 
\\
& & & 0 & & 0 & & &    & &
& & & 0 & & 0 & & & 
\\
& & & & 1 & & & &     & & 
& & & & LG$_{r=0,1,2}$, $\sigma$ & & & & 
\end{tabular}
\end{center}

\subsubsection{$n=6$}

In this subsection we consider ${\mathbb P}^6[4]$.
Mathematically, $D_p(n,4) = 546$, and
expanding the generating function, for fixed $n=6$, we have
\begin{equation}
\sum_{k,\ell} \left( h^{k,\ell} - \delta_{k,\ell} \right)
x^k y^{\ell}
\: = \:
7 x^4 y + 266 x^3 y^2 + 266 x^2 y^3 + 7 xy^4,
\end{equation}
from which we deduce
\begin{equation}
h^{4,1} = 7, \: \: \:
h^{3,2} = 266, \: \: \:
h^{2,3} = 266, \: \: \:
h^{1,4} = 7,
\end{equation}
plus
\begin{equation}
h^{0,0} \: = \: 1 \: = \: h^{1,1} \: = \: h^{2,2}
\: = \: h^{3,3} \: = \: h^{4,4} \: = \: h^{5,5}.
\end{equation}

Now, let us turn to the corresponding physics.
From the Coulomb branch, there are $\sigma$ fields, obeying the
relation~(\ref{eq:qc:ci})
\begin{equation}
\sigma^{7-1} \: = \: q (-4)^4 \sigma^3,
\end{equation}
or equivalently for our purposes
\begin{equation}
\sigma^3 \: = \: (-4)^4 q.
\end{equation}
The Coulomb branch only contributes three states to $H^{k,k}$.
We shall see the remainder arise as Landau-Ginzburg orbifold states.

Next, we consider the Landau-Ginzburg orbifold.
Here, for later use, note $\rho \in {\mathbb Z}_6$,
whose generator we label $\zeta$.  The states are as follows:
\begin{itemize}
\item $r=0$:  one state of charge $(q_L,q_R) = (0,0)$, $\rho$ eigenvalue $1$,
which contributes to $H^{k,\ell}$ with $\ell-k=0$ and
$k +\ell \equiv 0 \mod 6$, hence
a linear combination of $H^{0,0}$, $H^{3,3}$.
\item $r=1$:
states of the form
\begin{equation}
{\cal U}_{-1/2,1/2} \, f_p(x) | 0, {\rm RR} \rangle
\end{equation}
with $f_p \sim f_p + dW$, for $p = 1, 5, 9, 13$, satisfying~(\ref{eq:constr-p}).
\begin{itemize}
\item For $p=1$, possible $f_p$ (modulo the ideal) are of the form
\begin{equation}
x_i
\end{equation}
of which there are 7 possible terms.
The corresponding states have charge 
$(q_L,q_R) = (-13/4, +1/4)$, $\rho$ eigenvalue $\zeta^5$,
and contribute to $H^{k,\ell}$ for $\ell-k = -3$, $k+\ell \equiv 5 \mod 6$,
hence $H^{4,1}$.
\item For $p=5$, possible $f_p$ (modulo the ideal) are of the form
\begin{equation}
x_{i_1} x_{i_2} x_{i_3} x_{i_4} x_{i_5}, \: \: \:
x_{i_1} x_{i_2} x_{i_3} x_{i_4}^2, \: \: \:
x_{i_1} x_{i_2}^2 x_{i_3}^2,
\end{equation}
of which there are $21 + 140 + 105 = 266$ possible terms.
The corresponding states have
charge $(q_L,q_R) = (-9/4, +5/4)$, $\rho$ eigenvalue $\zeta^5$,
and contribute to $H^{k,\ell}$ for $\ell-k = -1$, $k+\ell \equiv 5 \mod 6$,
hence $H^{3,2}$.
\item For $p=9$, possible $f_p$ (modulo the ideal) are of the form
\begin{equation}
x_{i_1}^2 x_{i_2}^2 x_{i_3} \cdots x_{i_7}, \: \: \:
x_{i_1}^2 x_{i_2}^2 x_{i_3}^2 x_{i_4} x_{i_5} x_{i_6}, \: \: \:
x_{i_1}^2 x_{i_2}^2 x_{i_3}^2 x_{i_4}^2 x_{i_5},
\end{equation}
of which there are $21 + 140 + 105 = 266$ possible terms.
The corresponding terms have
charge $(q_L,q_R) = (-5/4, +9/4)$, $\rho$ eigenvalue $\zeta^5$,
and contribute to $H^{k,\ell}$ for $\ell-k = +1$, $k+\ell \equiv 5 \mod 6$,
hence $H^{2,3}$.
\item For $p=13$, possible $f_p$ (modulo the ideal) are of the form
\begin{equation}
x_{i_1}^2 x_{i_2}^2 x_{i_3}^2 x_{i_4}^2 x_{i_5}^2 x_{i_6}^2 x_{i_7},
\end{equation}
of which there are 7 possible terms.
The corresponding states have
charge $(q_L,q_R) = (-1/4, +13/4)$, $\rho$ eigenvalue $\zeta^5$,
and contribute to $H^{k,\ell}$ for $\ell-k = +3$, $k+\ell \equiv 5 \mod 6$,
hence $H^{1,4}$.
\end{itemize}
\item $r=2$: one state of charge $(q_L,q_R) = (-7/2, +7/2)$,
$\rho$ eigenvalue $\zeta^{10}$, which contributes to
$H^{k,\ell}$ for $\ell-k=0$, $k+\ell \equiv 10 \mod 6$,
hence a linear combination of $H^{2,2}$, $H^{5,5}$.
\item $r=3$: one state of charge $(q_L,q_R) = (-7/4, +7/4)$,
$\rho$ eigenvalue $\zeta^8$, which contributes to $H^{k,\ell}$ for
$\ell-k = 0$, $k+\ell \equiv 2 \mod 6$ hence a linear combination of
$H^{1,1}$, $H^{4,4}$.
\end{itemize}

The Hodge diamond for ${\mathbb P}^6[4]$ is
\begin{center}
\begin{tabular}{ccccccccccc}
& & & & & 1 & & & & & \\
& & & & 0 & & 0 & & & & \\
& & & 0 & & 1 & & 0 & & & \\
& & 0 & & 0 & & 0 & & 0 & & \\
& 0 & & 0 & & 1 & & 0 & & 0 \\
0 & & 7 & & 266 & & 266 & & 7 & & 0 \\
& 0 & & 0 & & 1 & & 0 & & 0 \\
& & 0 & & 0 & & 0 & & 0 & & \\
& & & 0 & & 1 & & 0 & & & \\
& & & & 0 & & 0 & & & & \\
& & & & & 1 & & & & & \\
\end{tabular}
\end{center}
We summarize the physical origin of these states below,
using $\sigma$ to indicate Coulomb-branch contributions from $\sigma$
fields, and LG to indicate Higgs-branch contributions from
Landau-Ginzburg-orbifold states:
\begin{center}
\begin{tabular}{ccccccccccc}
& & & & & LG$_{r=0}$, $\sigma$ & & & & & \\
& & & & 0 & & 0 & & & & \\
& & & 0 & & LG$_{r=3}$, $\sigma$ & & 0 & & & \\
& & 0 & & 0 & & 0 & & 0 & & \\
& 0 & & 0 & & LG$_{r=2}$, $\sigma$ & & 0 & & 0 \\
0 & & LG & & LG & & LG & & LG & & 0 \\
& 0 & & 0 & & LG$_{r=0}$, $\sigma$ & & 0 & & 0 \\
& & 0 & & 0 & & 0 & & 0 & & \\
& & & 0 & & LG$_{r=3}$, $\sigma$ & & 0 & & & \\
& & & & 0 & & 0 & & & & \\
& & & & & LG$_{r=2}$, $\sigma$ & & & & & \\
\end{tabular}
\end{center}

\subsubsection{$n=7$}

In this section we consider ${\mathbb P}^7[4]$.
Mathematically, $D_p(n,4) = 1641$, and
expanding the generating function, for fixed $n=7$, we have
\begin{equation}
\sum_{k,\ell} \left( h^{k,\ell} - \delta_{k,\ell} \right)
x^k y^{\ell}
\: = \:
x^5 y + 266 x^4 y^2 + 1107 x^3 y^3 + 266 x^2 y^4 + x y^5,
\end{equation}
from which we deduce
\begin{equation}
h^{5,1} = 1, \: \: \:
h^{4,2} = 266, \: \: \:
h^{3,3} = 1108, \: \: \:
h^{2,4} = 266, \: \: \:
h^{1,5} = 1,
\end{equation}
plus
\begin{equation}
h^{0,0} \: = \: 1 \: = \: h^{1,1} \: = \: h^{2,2}
\: = \: h^{4,4} \: = \: h^{5,5} \: = \: h^{6,6}.
\end{equation}

Now, let us turn to the corresponding physics.
From the Coulomb branch, there are $\sigma$ fields, obeying the
relation~(\ref{eq:qc:ci})
\begin{equation}
\sigma^{8-1} \: = \: q (-4)^4 \sigma^3,
\end{equation}
or equivalently for our purposes
\begin{equation}
\sigma^4 \: = \: (-4)^4 q.
\end{equation}
The Coulomb branch only contributes four states to $H^{k,k}$.
We will see the remainder arise as Landau-Ginzburg orbifold states.

Next, we consider the Landau-Ginzburg orbifold.
Here, for later use, note $\rho \in {\mathbb Z}_8$,
whose generator we label $\zeta$.  The states are as follows:
\begin{itemize}
\item $r=0$:  one state of charge $(q_L,q_R) = (0,0)$, $\rho$ eigenvalue $1$,
which contributes to $H^{k,\ell}$ for $\ell-k = 0$, $k + \ell \equiv 0 \mod 8$,
hence a linear combination of $H^{0,0}$, $H^{4,4}$.
\item $r=1$:
states of the form
\begin{equation}
{\cal U}_{-1/2,1/2} \, f_p(x) | 0, {\rm RR} \rangle
\end{equation}
with $f_p \sim f_p + dW$, for $p=0, 4, 8, 12, 16$, satisfying~(\ref{eq:constr-p}).
\begin{itemize}
\item For $p=0$, $f_p \propto 1$, hence one possible term.
The corresponding state has
charge $(q_L,q_R) = (-4, 0)$, $\rho$ eigenvalue $\zeta^6$,
and contributes to $H^{k,\ell}$ for $\ell-k = -4$, $k+\ell \equiv 6 \mod 8$,
hence $H^{5,1}$.
\item For $p=4$, possible $f_p$ (modulo the ideal) are of the form
\begin{equation}
x_{i_1} x_{i_2} x_{i_3} x_{i_4}, \: \: \:
x_{i_1}^2 x_{i_2} x_{i_3}, \: \: \:
x_{i_1}^2 x_{i_2}^2
\end{equation}
of which there are $70 + 168 + 28 = 266$ possible terms.
The corresponding states have 
charge $(q_L,q_R) = (-3,+1)$, $\rho$ eigenvalue $\zeta^6$,
and contribute to $H^{k,\ell}$ for $\ell-k = -2$, $k+\ell \equiv 6 \mod 8$,
hence $H^{4,2}$.
\item For $p=8$, possible $f_p$ (modulo the ideal) are of the form
\begin{equation}
\begin{array}{c}
x_{i_1} x_{i_2} x_{i_3} x_{i_4} x_{i_5} x_{i_6} x_{i_7} x_{i_8},
\: \: \:
x_{i_1}^2 x_{i_2} x_{i_3} x_{i_4} x_{i_5} x_{i_6} x_{i_7}, \: \: \:
x_{i_1}^2 x_{i_2}^2 x_{i_3} x_{i_4} x_{i_5} x_{i_6},
% \: \: \:
\\
x_{i_1}^2 x_{i_2}^2 x_{i_3}^2  x_{i_4} x_{i_5},
 \: \: \:
x_{i_1}^2 x_{i_2}^2 x_{i_3}^2 x_{i_4}^2,
\end{array}
\end{equation}
of which there are $1 + 56 + 420 + 560 + 70 = 1107$ possible terms.
The corresponding states have 
charge $(q_L,q_R) = (-2,+2)$, $\rho$ eigenvalue $\zeta^6$,
and contribute to $H^{k,\ell}$ for $\ell-k = 0$,
$k + \ell \equiv 6 \mod 8$, hence $H^{3,3}$.
\item For $p=12$, possible $f_p$ (modulo the ideal) are of the form
\begin{equation}
x_{i_1}^2 x_{i_2}^2 x_{i_3}^2 x_{i_4}^2 x_{i_5}^2 x_{i_6}^2, \: \: \:
x_{i_1}^2 x_{i_2}^2 x_{i_3}^2 x_{i_4}^2 x_{i_5}^2 x_{i_6} x_{i_7}, \: \: \:
x_{i_1}^2 x_{i_2}^2 x_{i_3}^2 x_{i_4}^2 x_{i_5} x_{i_6} x_{i_7} x_{i_8},
\end{equation}
of which there are $28 + 168 + 70 = 266$ possible terms.
The corresponding states have
charge $(q_L,q_R) = (-1,+3)$, $\rho$ eigenvalue $\zeta^6$,
and contribute to $H^{k,\ell}$ for $\ell-k = +2$,
$k+\ell \equiv 6 \mod 8$, hence $H^{2,4}$.
\item For $p=16$, possible $f_p$ (modulo the ideal) are of the form
\begin{equation}
x_{i_1}^2 x_{i_2}^2 x_{i_3}^2 x_{i_4}^2 x_{i_5}^2 x_{i_6}^2 x_{i_7}^2 x_{i_8}^2,
\end{equation}
of which there is only one term.
The corresponding state has
charge $(q_L,q_R) = (0,+4)$, $\rho$ eigenvalue $\zeta^6$,
and contributes to $H^{k,\ell}$ for $\ell-k = +4$, $k+\ell \equiv 6 \mod 8$,
hence $H^{1,5}$.
\end{itemize}
\item $r=2$: one state of charge $(q_L,q_R) = (-4,+4)$,
$\rho$ eigenvalue $\zeta^{12}$, which contributes to
$H^{k,\ell}$ for $\ell-k=0$, $k+\ell \equiv 12 \mod 8$ hence
a linear combination of $H^{2,2}$ and $H^{6,6}$.
\item $r=3$: one state of charge $(q_L,q_R) = (-2,+2)$, $\rho$ eigenvalue
$\zeta^{10}$, which contributes to $H^{k,\ell}$ for $ell-k=0$,
$k+\ell \equiv 10 \mod 8$, hence a linear combination of 
$H^{1,1}$ and $H^{5,5}$.
\end{itemize}

The Hodge diamond for ${\mathbb P}^7[4]$ is
\begin{center}
\begin{tabular}{ccccccccccccc}
& & & & & & 1 & & & & & & \\
& & & & & 0 & & 0 & & & & & \\
& & & & 0 & & 1 & & 0 & & & & \\
& & & 0 & & 0 & & 0 & & 0 & & & \\
& & 0 & & 0 & & 1 & & 0 & & 0 & & \\
& 0 & & 0 & & 0 & & 0 & & 0 & & 0 & \\
0 & & 1 & & 266 & & 1108 & & 266 & & 1 & & 0 \\
& 0 & & 0 & & 0 & & 0 & & 0 & & 0 & \\
& & 0 & & 0 & & 1 & & 0 & & 0 & & \\
& & & 0 & & 0 & & 0 & & 0 & & & \\
& & & & 0 & & 1 & & 0 & & & & \\
& & & & & 0 & & 0 & & & & & \\
& & & & & & 1 & & & & & & 
\end{tabular}
\end{center}
We summarize the physical origin of these states below,
using $\sigma$ to indicate Coulomb-branch contributions from $\sigma$
fields, and LG to indicate Higgs-branch contributions from
Landau-Ginzburg-orbifold states:
\begin{center}
\begin{tabular}{ccccccccccccc}
& & & & & & LG$_{r=0}$, $\sigma$ & & & & & & \\
& & & & & 0 & & 0 & & & & & \\
& & & & 0 & & LG$_{r=3}$, $\sigma$ & & 0 & & & & \\
& & & 0 & & 0 & & 0 & & 0 & & & \\
& & 0 & & 0 & & LG$_{r=2}$, $\sigma$ & & 0 & & 0 & & \\
& 0 & & 0 & & 0 & & 0 & & 0 & & 0 & \\
0 & & LG & & LG & & LG$_{r=1}$, $\sigma$ & & LG & & LG & & 0 \\
& 0 & & 0 & & 0 & & 0 & & 0 & & 0 & \\
& & 0 & & 0 & & LG$_{r=0}$, $\sigma$ & & 0 & & 0 & & \\
& & & 0 & & 0 & & 0 & & 0 & & & \\
& & & & 0 & & LG$_{r=3}$, $\sigma$ & & 0 & & & & \\
& & & & & 0 & & 0 & & & & & \\
& & & & & & LG$_{r=2}$, $\sigma$ & & & & & & 
\end{tabular}
\end{center}

\subsubsection{$n=8$}

In this section we consider ${\mathbb P}^8[4]$.
Mathematically, $D_p(n,4) = 4920$, and
expanding the generating function, for fixed $n=8$, we have
\begin{equation}
\sum_{k,\ell} \left( h^{k,\ell} - \delta_{k,\ell} \right)
x^k y^{\ell}
\: = \:
156 x^5 y^2 + 2304 x^4 y^3 + 2304 x^3 y^4 + 156 x^2 y^5,
\end{equation}
from which we deduce
\begin{equation}
h^{5,2} = 156, \: \: \:
h^{4,3} = 2304, \: \: \:
h^{3,4} = 2304, \: \: \:
h^{2,5} = 156,
\end{equation}
plus
\begin{equation}
h^{0,0} \: = \: 1 \: = \: h^{1,1} \: = \: h^{2,2}
\: = \: h^{3,3} \: = \: h^{4,4} \: = \: h^{5,5} \: = \: h^{6,6} \: = \:
h^{7,7}.
\end{equation}

Now, let us turn to the corresponding physics.
From the Coulomb branch, there are $\sigma$ fields, obeying the
relation~(\ref{eq:qc:ci})
\begin{equation}
\sigma^{9-1} \: = \: q (-4)^4 \sigma^3,
\end{equation}
or equivalently for our purposes
\begin{equation}
\sigma^5 \: = \: (-4)^4 q.
\end{equation}
The Coulomb branch only contributes five states to $H^{k,k}$.
We shall see that the remainder arise as Landau-Ginzburg orbifold states.

Next, we consider the Landau-Ginzburg orbifold.
Here, for later use, note $\rho \in {\mathbb Z}_{10}$,
whose generator we label $\zeta$.  The states are as follows:
\begin{itemize}
\item $r=0$:  one state of charge $(q_L,q_R) = (0,0)$, $\rho$ eigenvalue $1$,
which contributes to $H^{k,\ell}$ for $\ell-k = 0$, $k+\ell \equiv 0 \mod 10$,
hence a linear combination of $H^{0,0}$, $H^{5,5}$.
\item $r=1$:
states of the form
\begin{equation}
{\cal U}_{-1/2,1/2} \, f_p(x) | 0, {\rm RR} \rangle
\end{equation}
with $f_p \sim f_p + dW$, for $p=3, 7, 11, 15$, satisfying~(\ref{eq:constr-p}).
\begin{itemize}
\item For $p=3$, possible $f_p$ (modulo the ideal) are of the form
\begin{equation}
x_{i_1}^2 x_{i_2}, \: \: \:
x_{i_1} x_{i_2} x_{i_3}
\end{equation}
of which there are $72 + 84 = 156$ possible terms.
The corresponding states have 
charge $(q_L,q_R) = (-15/4, +3/4)$, $\rho$ eigenvalue $\zeta^7$,
and contribute to $H^{k,\ell}$ for $\ell-k=-3$,
$k+\ell \equiv 7 \mod 10$, hence $H^{5,2}$.
\item For $p=7$, possible $f_p$ (modulo the ideal) are of the form
\begin{equation}
x_{i_1}^2 x_{i_2}^2 x_{i_3}^2 x_{i_4}, \: \: \:
x_{i_1}^2 x_{i_2}^2 x_{i_3} x_{i_4} x_{i_5}, \: \: \:
x_{i_1}^2 x_{i_2} x_{i_3} x_{i_4} x_{i_5} x_{i_6}, \: \: \:
x_{i_1} x_{i_2} x_{i_3} x_{i_4} x_{i_5} x_{i_6} x_{i_7},
\end{equation}
of which there are $504 + 1260 + 504 + 36 = 2304$ possible terms.
The corresponding states have
charge $(q_L,q_R) = (-11/4, +7/4)$, $\rho$ eigenvalue $\zeta^7$,
and contribute to $H^{k,\ell}$ for $\ell-k = -1$, $k+\ell \equiv 7 \mod 10$,
hence $H^{4,3}$.
\item For $p=11$, possible $f_p$ (modulo the ideal) are of the form
\begin{equation}
\begin{array}{c}
x_{i_1}^2 x_{i_2}^2 x_{i_3}^2 x_{i_4}^2 x_{i_5}^2 x_{i_6}, \: \: \:
x_{i_1}^2 x_{i_2}^2 x_{i_3}^2 x_{i_4}^2 x_{i_5} x_{i_6} x_{i_7},
% \: \: \:
\\
x_{i_1}^2 x_{i_2}^2 x_{i_3}^2 x_{i_4} x_{i_5} x_{i_6} x_{i_7} x_{i_8},
\: \: \:
x_{i_1}^2 x_{i_2}^2 x_{i_3} x_{i_4} x_{i_5} x_{i_6} x_{i_7} x_{i_8} x_{i_9},
\end{array}
\end{equation}
of which there are $504 + 1260 + 504 + 36 = 2304$ possible terms.
The corresponding states have
charge $(q_L,q_R) = (-7/4, +11/4)$, $\rho$ eigenvalue $\zeta^7$,
and contribute to $H^{k,\ell}$ for $\ell-k = +1$, $k+\ell \equiv 7 \mod 10$,
hence $H^{3,4}$.
\item For $p=15$, possible $f_p$ (modulo the ideal) are of the form
\begin{equation}
x_{i_1}^2 x_{i_2}^2 x_{i_3}^2 x_{i_4}^2 x_{i_5}^2 x_{i_6}^2 x_{i_7}^2 x_{i_8},
\: \: \:
x_{i_1}^2 x_{i_2}^2 x_{i_3}^2 x_{i_4}^2 x_{i_5}^2 x_{i_6}^2 x_{i_7} x_{i_8} x_{i_9},
\end{equation}
of which there are $72 + 84 = 156$ possible terms.
The corresponding states have
charge $(q_L,q_R) = (-3/4, +15/4)$, $\rho$ eigenvalue $\zeta^7$,
and contribute to $H^{k,\ell}$ for $\ell-k = +3$, $k+\ell \equiv 7 \mod 10$,
hence $H^{2,5}$.
\end{itemize}
\item $r=2$: one state of charge $(q_L,q_R) = (-9/2, +9/2)$,
$\rho$ eigenvalue $\zeta^{14}$, which contributes to $H^{k,\ell}$
for $\ell-k=0$, $k+\ell \equiv 14 \mod 10$, hence a linear combination
of $H^{2,2}$, $H^{7,7}$.
\item $r=3$: one state of charge $(q_L,q_R) = (-9/4, +9/4)$, $\rho$
eigenvalue $\zeta^{12}$, which contributes to $H^{k,\ell}$ for
$\ell-k =0$, $k+\ell \equiv 12 \mod 10$, hence
a linear combination of $H^{1,1}$, $H^{6,6}$.
\end{itemize}

The Hodge diamond for ${\mathbb P}^8[4]$ is
\begin{center}
\begin{tabular}{ccccccccccccccc}
& & & & & & & 1 & & & & & & & \\
& & & & & & 0 & & 0 & & & & & & \\
& & & & & 0 & & 1 & & 0 & & & & & \\
& & & & 0 & & 0 & & 0 & & 0 & & & & \\
& & & 0 & & 0 & & 1 & & 0 & & 0 & & & \\
& & 0 & & 0 & & 0 & & 0 & & 0 & & 0 & & \\
& 0 & & 0 & & 0 & &  1 & & 0 & & 0 & & 0 & \\
0 & & 0 & & 156 & & 2304 & & 2304 & & 156 & & 0 & & 0 \\
& 0 & & 0 & & 0 & &  1 & & 0 & & 0 & & 0 & \\
& & 0 & & 0 & & 0 & & 0 & & 0 & & 0 & & \\
& & & 0 & & 0 & & 1 & & 0 & & 0 & & & \\
& & & & 0 & & 0 & & 0 & & 0 & & & & \\
& & & & & 0 & & 1 & & 0 & & & & & \\
& & & & & & 0 & & 0 & & & & & & \\
& & & & & & & 1 & & & & & & &
\end{tabular}
\end{center}
We summarize the physical origin of these states below,
using $\sigma$ to indicate Coulomb-branch contributions from $\sigma$
fields, and LG to indicate Higgs-branch contributions from
Landau-Ginzburg-orbifold states:
\begin{center}
\begin{tabular}{ccccccccccccccc}
& & & & & & & LG$_{r=0}$, $\sigma$ & & & & & & & \\
& & & & & & 0 & & 0 & & & & & & \\
& & & & & 0 & & LG$_{r=3}$, $\sigma$ & & 0 & & & & & \\
& & & & 0 & & 0 & & 0 & & 0 & & & & \\
& & & 0 & & 0 & & LG$_{r=2}$, $\sigma$ & & 0 & & 0 & & & \\
& & 0 & & 0 & & 0 & & 0 & & 0 & & 0 & & \\
& 0 & & 0 & & 0 & &  $\sigma$ & & 0 & & 0 & & 0 & \\
0 & & 0 & & LG & & LG & & LG & & LG & & 0 & & 0 \\
& 0 & & 0 & & 0 & &  $\sigma$ & & 0 & & 0 & & 0 & \\
& & 0 & & 0 & & 0 & & 0 & & 0 & & 0 & & \\
& & & 0 & & 0 & & LG$_{r=0}$, $\sigma$ & & 0 & & 0 & & & \\
& & & & 0 & & 0 & & 0 & & 0 & & & & \\
& & & & & 0 & & LG$_{r=3}$, $\sigma$ & & 0 & & & & & \\
& & & & & & 0 & & 0 & & & & & & \\
& & & & & & & LG$_{r=2}$, $\sigma$ & & & & & & &
\end{tabular}
\end{center}

\section{Other theories}
\label{sect:other}

In this section we will apply the same ideas to computations in
quantum K theory (realized via three-dimensional GLSMs)
and quantum sheaf cohomology (realized via two-dimensional GLSMs with
(0,2) supersymmetry).  We will see that the structure of the mathematical
results is consistent with the form of the computations described earlier;
unfortunately, in both cases there is insufficient symmetry to make 
a prediction just from the physics we currently understand.

\subsection{Quantum K theory}

Quantum K theory can be computed from three-dimensional GLSMs,
see e.g.~\cite{Bullimore:2014awa,Jockers:2018sfl,Jockers:2019wjh,Jockers:2019lwe,Ueda:2019qhg,Gu:2020zpg,Gu:2022yvj,Gu:2023tcv}.
Consider a three-manifold of the form $S^1 \times \Sigma$,
for a Riemann surface $\Sigma$.  The quantum K theory relations emerge
as the OPEs of Wilson lines wrapped on $S^1$ as they are brought closer
along $\Sigma$.  By performing a Kaluza-Klein reduction to $\Sigma$
(and resumming, not truncating, the higher modes), one recovers an
effective two-dimensional theory in which OPEs of local operators
encode the Wilson line OPEs of the three-dimensional theory.

For GLSMs without superpotential
describing Fano target spaces, these ideas have been used
to compute quantum K theory ring relations using Coulomb
branch relations, in the same pattern as \cite{Morrison:1994fr}.
It is natural to try to extend these computations to
Fano spaces described by
GLSMs with superpotential, using the methods described earlier.

Unfortunately,
for quantum K theory, in general
we are only able to use the physical methods above
to make weaker statements, essentially because the observables in the quantum
K theory ring do not have well-defined $U(1)_R$
eigenvalues, so internal consistency is a weaker
constraint.  We describe some computations below.

For ${\mathbb P}^n[d]$, the quantum K theory ring
determined solely by the Coulomb branch ($\sigma$ fields) has the
form
\cite[equ'n (5.7)]{Gu:2020zpg},
\cite[equ'n (2.24)]{Jockers:2018sfl}, namely
\begin{equation}  \label{eq:qk:hyp}
(1-x)^{n+1} \: = \: (-)^d q (1-x^d)^d,
\end{equation}
and we also assume~(\ref{eq:root}) that $\sigma \cdot t = 0$, or more explicitly
$(1-x) t = 0$, for any Landau-Ginzburg orbifold state.
(we use here $\sigma \sim 1-x$.)
This will determine some, but not all, of the quantum K theory
ring relations.

Let us briefly walk through two examples.

First, consider the quantum K theory of a hyperplane,
${\mathbb P}^n[1]$.  For the same reasons discussed earlier in
section~\ref{sect:exs:deg1} and in \cite[section 5.2]{Gu:2020zpg},
the superpotential acts as a mass term, removing both $p$ and $x_{n+1}$,
reducing this theory to the ${\mathbb P}^{n-1}$ model,
with quantum K theory given by that of ${\mathbb P}^{n-1}$.

Next, we consider degree two hypersurfaces.
Our analysis in this case closely follows that of quantum cohomology
for degree two hypersurfaces.  In the Landau-Ginzburg phase, for a nondegenerate
quadric $Q$, the $x_i$ are massive, and there are $n+1$ of them.

Just as in our discussion of quantum cohomology of quadrics,
if $n+1$ is odd, taking the ${\mathbb Z}_2$ orbifold results in a single
vacuum, whereas if $n+1$ is even, taking the ${\mathbb Z}_2$ orbifold
results in a pair of vacua.
Specializing~(\ref{eq:qk:hyp}), the $\sigma$ field contribution obeys
\begin{equation}
(1-x)^{n+1} \: = \: q \left(1-x^2\right)^2,
\end{equation}

For $n+1$ odd, as there are no nontrivial Landau-Ginzburg orbifold vacua,
one expects that this is the complete quantum K theory ring of
${\mathbb P}^n[2]$.  Indeed, for $n+1$ odd, this matches known 
results for this case (see e.g.~\cite{Gu:2020zpg} and references therein),
under the dictionary
$1-x = {\cal O}_{\tiny\yng(1)}$.

For $n+1$ even, for the same reason as in quantum cohomology
computations, physics predicts that the quantum K theory ring
of ${\mathbb P}^n[2]$
has an extra generator $t$ in (middle) degree $k = n-1$ and
relations
\begin{eqnarray}
(1-x)^{n+1} & = & q \left(1-x^2\right)^2,
\\
(1-x) \cdot t & = & 0,
\end{eqnarray}
using~(\ref{eq:root}), plus a relation involving $t^2$, to which we turn
next.

To get the $t^2$ relation,
we can try to follow the same procedure as for quantum cohomology.
Write $y = 1-x$.  Since the pertinent operators do not have
well-defined $U(1)_R$ eigenvalues,
we can only say that $t^2$ is some polynomial in $y$.
Applying $y t^2 = 0$, requires that
\begin{equation}
yt^2 \: \propto \:
f(y) \left( y^n - q y (y-2)^2 \right),
\end{equation}
for some unknown function $f(y)$
We can then read off that
\begin{equation}
t^2 \: = \: f(y) \left( y^{n-1} - q (y-2)^2 \right)
\end{equation}
(up to a proportionality factor which can be absorbed into the definition
of $t$).  

To completely determine the relation, we would need
to determine the function $f(y)$.
In quantum cohomology, $f(y)$ was determined on symmetry grounds,
but that is not an option here.
We leave the physical determination of $f(y)$ for future work.

We can compare to existing mathematics results\footnote{
We would like to thank L.~Mihalcea and W.~Xu for discussions of
the quantum K theory ring of ${\mathbb P}^n[2]$.
} for the case $n+1$
even, which determine $f(y)$ to be a constant.
Specifically a presentation of the quantum K theory ring of
${\mathbb P}^n[2]$ for $n+1$ even is given by
\begin{equation}
{\mathbb C}[y,t,q] / \langle
y^{2k+1} - q y (y-2)^2, \: \: \:
yt = 0, \: \: \:
t^2 = (-1)^k (y^{2k} - q(y-2)^2)
\rangle
\end{equation}
under the dictionary
\begin{equation}
y \: = \: 1-x \: = \: {\cal O}_{\tiny\yng(1)},
\: \: \:
t \: = \: {\cal O}_{k} \: - \: {\cal O}_{k-1,1}.
\end{equation}
Here the notation ${\cal O}$ indicates a Schubert class
\begin{equation}
{\cal O}_0, \: \: \:
{\cal O}_1 = {\cal O}_{\tiny\yng(1)}, \: \: \:
\cdots, \: \: \:
 {\cal O}_{k}, \: \: \:
{\cal O}_{k-1,1}, \: \: \:
{\cal O}_{k,1}, \: \: \:
{\cal O}_{k,k},
\end{equation}
where $n=2k+1$.
Linearizing this result reproduces the
quantum cohomology ring~(\ref{eq:qh:evendiml-quadric}) 
of the quadric ${\mathbb P}^n[2]$ for $n$ odd, 
namely
\begin{equation} 
{\mathbb C}[y,t,q] / \langle
y^{2k+1} - 4qy, \: \: \:
yt = 0, \: \: \:
t^2 = (-1)^k( y^{2k} - 4q )
\rangle
\end{equation}
where $n = 2k+1$, $y=1-x$.

These matters have also been considered in \cite{Gu:2021yek},
where further subtleties are discussed.

\subsection{Quantum sheaf cohomology}

Quantum sheaf cohomology \cite{Katz:2004nn} is a 
quantum-corrected sheaf cohomology ring, realized physically
via twisted two-dimensional (0,2) supersymmetric theories.
It can also be realized in gauged linear sigma models.
For Fano toric varieties, it was computed using GLSM techniques
in \cite{McOrist:2007kp,McOrist:2008ji},
and mathematically in \cite{Donagi:2011va,Donagi:2011uz}.
See also \cite{Guo:2015caf,Guo:2016suk} for computations on Grassmannians,
and \cite{Guo:2018iyr} for computations on flag manifolds.
See also \cite{McOrist:2010ae,Guffin:2011mx,Melnikov:2012hk} and
references therein for reviews and additional references.

For quantum sheaf cohomology, 
we expect an analogue of~(\ref{eq:root}),
the statement that $\sigma t = 0$ in OPE rings, essentially because of the
(0,2) GLSM bosonic potential term $|E(\phi)|^2$,
where on the (2,2) locus,
\begin{equation}
E(\phi_i) \: = \: \sigma Q_i \phi_i.
\end{equation}
Proceeding as before, one expects that the OPE ring should include the
product $\sigma t = 0$.

That said, the methods we have described so far do not suffice
to describe the quantum cohomology ring of a (0,2) deformation of a 
hypersurface.
The reason\footnote{
Well, one reason.  As is well-known, for (2,2) cases, the dimension of the chiral ring is
invariant under deformations.  For (0,2) theories, matters are slightly
more subtle.  For example, in geometric realizations, the states correspond
to sheaf cohomology, which can jump due to the semicontinuity theorem,
see e.g.~\cite[section III.12]{hartshorne}.
One particularly vivid recent example of this subtlety in (0,2) theories
along RG flows specifically is described
in \cite{Guo:2015gha}.  In any event, our central concerns here are 
different in nature.
} is that ultimately (0,2) deformations of hypersurfaces
are encoded in $J$ deformations, whereas typical quantum sheaf cohomology
computations have only been worked out for $E$ deformations, as we shall
discuss.  Now, there is a duality that suggests one may be able to make
progress:  the Grassmannian $G(2,4) = {\mathbb P}^5[2]$, and (0,2) deformations
of $G(2,4)$ are $E$-deformations and hence computable.  Unfortunately, we 
shall see that the dictionary to ${\mathbb P}^5[2]$ is sufficiently
obscure to make any possible conclusions about quantum sheaf
cohomology of $J$ deformations of ${\mathbb P}^5[2]$, much less 
hypersurfaces of other degrees, unreachable at this time.
For completeness, we will quickly walk through both of these points.

\subsubsection{Basic setup}

In (0,2) language, in a $U(1)$ theory with
chiral fields $\Phi$ of charge $Q$,
\begin{equation}
E \: = \: Q \sigma \Phi, \: \: \:
J \: = \: \frac{\partial W}{\partial \Phi}.
\end{equation}
on the (2,2) locus.  
A quadric in ${\mathbb P}^n$ is described in (0,2) language on the (2,2)
locus as
\begin{equation}
\begin{array}{cc}
E_p \: = \: -2 \sigma p, & J_p \: = \: Q(x_i),
\\
E_i \: = \: \sigma x_i, & J_i \: = \: p \, \partial_i Q(x_i).
\end{array}
\end{equation}
We can deform off the (2,2) locus by adding terms to the $J$'s,
such that
\begin{equation}  \label{eq:ej-eq-zero}
\sum_{{\rm field} \: \alpha} E_{\alpha} J_{\alpha} \: = \: 0.
\end{equation}
In the present case, for (0,2) deformations of ${\mathbb P}^n[2]$,
this can be accomplished by taking
\begin{equation}
\begin{array}{cc}
E_p \: = \: -2 \sigma p, & J_p \: = \: Q(x_i),
\\
E_i \: = \: \sigma x_i, & J_i \: = \: p \, \partial_i Q(x_i) \: + \:
\sum_j p A_{ij} x_j,
\end{array}
\end{equation}
where $A_{ij}$ is an antisymmetric $(n+1)\times(n+1)$ matrix.
It is straightforward to check that this satisfies~(\ref{eq:ej-eq-zero}).

We can simplify the expression above by observing that
\begin{equation}
\partial_i Q \: = \: \sum_j S_{ij} x_j
\end{equation}
for a symmetric matrix $S_{ij}$, hence
\begin{equation}
J_i \: = \: p \sum_j M_{ij} x_j,
\end{equation}
where $M_{ij}$ is a general $(n+1)\times(n+1)$ matrix whose symmetric
part is determined by derivatives of $Q$.

In any event, we now see the basic problem:  current techniques for
computing quantum sheaf cohomology are restricted to $E$ deformations,
but the (0,2) deformations of a hypersurface in ${\mathbb P}^n$ are
instead deformations of $J$, for which no computational methods currently
exist.

In the next subsection, we will discuss a possible workaround in 
a special case, for which
a duality exists to a different description, in which the
(0,2) deformations are realized by $E$ deformations, not $J$ deformations,
and so quantum sheaf cohomology is computable.

\subsubsection{Special case: ${\mathbb P}^5[2] = G(2,4)$}

One family of cases for which we have results are (0,2) deformations of
${\mathbb P}^5[2]$, using the fact that ${\mathbb P}^5[2]$ is the
same as the Grassmannian $G(2,4)$, and results on quantum sheaf cohomology
for Grassmannians are known \cite{Guo:2015caf,Guo:2016suk}.
We shall rewrite known results for $G(2,4)$ in the form of results for
the quadric ${\mathbb P}^5[2]$.  We shall see that when expressed
in the language of the presentation ${\mathbb P}^5[2]$,
the quantum sheaf cohomology ring can be described in terms of
generators $\sigma$, $t$ (on the Coulomb and Higgs branches, respectively),
such that $\sigma \cdot t = 0$, as expected for a mixed branch OPE.
Unfortunately, the form of $t$ is obscure, and so we are not able to use
this as a key for predicting results for more general hypersurfaces.

Before describing (0,2) deformations, let us first quickly review
the (2,2) supersymmetric $G(2,4)$ and its (ordinary) quantum cohomology ring.
The cohomology ring of any Grassmannian $G(k,n)$ is described additively
by Young tableau fitting inside a $k \times n$ box, which for $G(2,4)$
means the cohomology ring has the additive generators
\begin{equation}
1, \: \: \:
\sigma_{(1)} = \sigma_{\tiny\yng(1)}, \: \: \:
\sigma_{(2)} = \sigma_{\tiny\yng(2)}, \: \: \:
\sigma_{\tiny\yng(1,1)}, \: \: \:
\sigma_{\tiny\yng(2,1)}, \: \: \:
\sigma_{\tiny\yng(2,2)},
\end{equation}
with cohomological degree given by twice the number of boxes.
Multiplicatively, this ring has relations
\begin{eqnarray}
\sigma_{\tiny\yng(1)} \sigma_{\tiny\yng(2)} 
& = &
\sigma_{\tiny\yng(1)} \sigma_{\tiny\yng(1,1)},
\label{eq:g24:22:1}
\\
\sigma_{\tiny\yng(1,1)} & = & \sigma_{(1)}^2 - \sigma_{(2)},
\\
\sigma_{\tiny\yng(2,1)} & = & \sigma_{(1)} \sigma_{\tiny\yng(1,1)},
\\
\sigma_{\tiny\yng(2,2)} & = & \sigma_{(2)}^2 - \sigma_{(1)}^2 \sigma_{(2)}
+ \sigma_{(1)}^2 \sigma_{\tiny\yng(1,1)},
\\
& = &  \sigma_{\tiny\yng(1,1)}^2,
\end{eqnarray}
\begin{equation}
\sigma_{\tiny\yng(1)}^4 \: + \: \sigma_{\tiny\yng(1)}^2 \sigma_{\tiny\yng(2)}
\: + \:
\sigma_{\tiny\yng(2)}^2 \: - \: \sigma_{\tiny\yng(1)}^2 \sigma_{\tiny\yng(1,1)}
\: = \: -q.
\end{equation}
Note that there are two degree four elements, $\sigma_{\tiny\yng(2)}$ and
$\sigma_{\tiny\yng(1,1)}$,  and as we now argue, the nontrivial Landau-Ginzburg field $t$
arising in the presentation ${\mathbb P}^5[2]$
is a linear combination of these elements.
It is natural to identify the ${\mathbb Z}_2$ quantum symmetry of the
Landau-Ginzburg orbifold with 
a transpose operation on the Young tableau, which leaves elements of
all other degrees invariant and exchanges $\sigma_{\tiny\yng(2)}$ and $\sigma_{\tiny\yng(1,1)}$.
Up to scalar multiples, we then identify
\begin{equation}
t \: = \: \sigma_{\tiny\yng(2)} \: - \: \sigma_{\tiny\yng(1,1)}~.
\end{equation}
As there is only one generator of
cohomological degree 2, namely $\sigma_{\tiny\yng(1)}$,
we identify, up to scalar multiplication,
\begin{equation}
\sigma \: = \: \sigma_{\tiny\yng(1)}.
\end{equation}
The reader should note that
equation~(\ref{eq:g24:22:1}) then implies equation~(\ref{eq:root}),
namely,
\begin{equation}
\sigma \cdot t \: = \: 0.
\end{equation}

Now, (0,2) deformations of a Grassmannian $G(k,n)$ are defined by
a traceless $n \times n$ matrix $B$, as described in
\cite{Guo:2015caf,Guo:2016suk}.
The resulting quantum sheaf cohomology ring is defined in terms of
functions $I_j$ defined
 \cite[section 2.2]{Guo:2015caf} to be
the coefficients of the characteristic
polynomial of $B$:
\begin{equation}
\det (t I + B) \: = \: \sum_{i=0}^{n} I_{n-i} t^i.
\end{equation}
For example,
\begin{equation}
I_0 \: = \: 1, \: \: \:
I_1 \: = \: {\rm tr}\, B, \: \: \:
I_n \: = \: \det \, B.
\end{equation}

For the special case of (0,2) deformations of $G(2,4)$,
the quantum sheaf cohomology ring computed in
\cite{Guo:2015caf} can be described in terms of
the additive generators
\begin{equation}
1, \: \: \:
\sigma_{(1)}, \: \: \:
\sigma_{(2)}, \: \: \:
\sigma_{\tiny\yng(1,1)}, \: \: \:
\sigma_{\tiny\yng(2,1)}, \: \: \:
\sigma_{\tiny\yng(2,2)},
\end{equation}
by the relations
\begin{equation}  \label{eq:g24:1}
\sigma_{(1)} \sigma_{(2)} \left[ 1 + I_1 + I_2 + I_3 \right]
\: = \:
\sigma_{(1)} \sigma_{\tiny\yng(1,1)} \left[ 1 - I_2 - I_3 \right],
\end{equation}
\begin{eqnarray}
\sigma_{\tiny\yng(1,1)} & = & \sigma_{(1)}^2 - \sigma_{(2)},
\\
\sigma_{\tiny\yng(2,1)} & = & \sigma_{(1)} \sigma_{\tiny\yng(1,1)},
\\
\sigma_{\tiny\yng(2,2)} & = & \sigma_{(2)}^2 - \sigma_{(1)}^2 \sigma_{(2)}
+ \sigma_{(1)}^2 \sigma_{\tiny\yng(1,1)},
\\
& = &  \sigma_{\tiny\yng(1,1)}^2,
\end{eqnarray}
\begin{equation}
\sigma_{(1)}^4\left[ 1 + I_3 + I_4 \right] +
\sigma_{(1)}^2 \sigma_{(2)} \left[ -1 + I_1 + I_2 \right]
+ \sigma_{(2)}^2 - \sigma_{(1)}^2 \sigma_{\tiny\yng(1,1)} \left[ 2 + I_1 
\right]
\: = \: -q.
\end{equation}

As a consistency check,
on the (2,2) locus, the relations above reduce to
\begin{equation}
\begin{array}{c}
\sigma_{(1)} \sigma_{(2)} \: = \: \sigma_{(1)} \sigma_{\tiny\yng(1,1)}
\: = \: \sigma_{\tiny\yng(2,1)}~,
\\
\sigma_{\tiny\yng(2,2)} \: = \: \sigma_{(2)}^2 \: = \: 
\sigma_{\tiny\yng(1,1)}^2~,
\\
\sigma_{(2)} \sigma_{\tiny\yng(1,1)} \: = \: q~,
\\
\sigma_{(1)} \sigma_{\tiny\yng(2,1)} \: = \: 
\sigma_{\tiny\yng(2,2)} + q~,
\end{array}
\end{equation}
which is a presentation of the ordinary quantum cohomology ring of
$G(2,4)$.

If we naively identify
\begin{equation}
t \: = \: \sigma_{(2)} \: - \: \sigma_{\tiny\yng(1,1)}~,
\end{equation}
as we did on the (2,2) locus,
then we see from~(\ref{eq:g24:1}) that $\sigma t \neq 0$.
However, there is no fundamental reason why $t$ should be identified
in the same fashion as on the (2,2) locus.
Instead, it is also clear from~(\ref{eq:g24:1}) that if we instead pick
\begin{equation} \label{eq:g24:02:t}
t \: = \: \sigma_{(2)} \left[ 1 + I_1 + I_2 + I_3 \right]
\: - \:
\sigma_{\tiny\yng(1,1)} \left[ 1 - I_2 - I_3 \right],
\end{equation}
then $\sigma \cdot t = 0$, as expected.

We believe that the correct identification with the (0,2) deformation
of ${\mathbb P}^5[2]$ is the relation~(\ref{eq:g24:02:t}) above,
which is consistent with expectations.
Clearly, renormalization group flow is being shifted by the (0,2)
deformations in such a way as to make the identification of $t$ somewhat
obscure.
Unfortunately, we do not know of a way to identify $t$ from first principles,
and so the arguments of this paper are not sufficient to make predictions
for quantum sheaf cohomology in mixed Higgs-Coulomb phases.

\section{Conclusions}

In this paper, we have generalized the Coulomb-branch based
computations of quantum cohomology described in \cite{Morrison:1994fr}
to cases with mixed Higgs-Coulomb branches, discussing in detail
the case of GLSMs for hypersurfaces in projective spaces,
${\mathbb P}^n[d]$.
We have described in detail how the vector space and product structures
are reproduced by physical computations in the IR phase with both Higgs and Coulomb branches.

One of the conclusions of this work is that the cohomology of a Fano hypersurface
has a decomposition corresponding to the Coulomb and Higgs sectors that
is distinct from the decomposition of Dolbeault cohomology into primitive and
non-primitive subspaces.  It may be interesting to explore the mathematical
implications of this Higgs/Coulomb decomposition further.

In terms of generalizing our results, perhaps the most immediate next step is to consider GLSMs for complete intersection Fano varieties in $\P^n$.  In this case the Higgs sector of the IR phase is a hybrid theory, and a detailed comparison of the well-understood UV phase may clarify a number of aspects of hybrid CFTs.  As these are perhaps the closest to a ``generic'' Higgs branch of a GLSM, it would be quite useful to have more detailed studies of their structure in a controlled setting.

It would also be interesting and reasonably straightforward to repeat the correlation function
computations of section~\ref{sect:phys} using the technology
of supersymmetric localization, as in e.g.~\cite{Closset:2015rna}, as well as to generalize
our results to hypersurfaces in (resolved) weighted projective spaces.   It should also be instructive to consider higher rank GLSMs, where the Higgs/Coulomb decomposition will
become more intricate and there will be genuine mixed sectors, where some $\sigma$ fields have
large expectation values, while others are set to zero.  

Finally, it remains an open problem to extend these ideas to quantum K theory and
quantum sheaf cohomology computations for Fano spaces described
by GLSMs with superpotential.  In both cases, as we have seen,
the ideas we have
described in this paper are not sufficient to completely predict
the OPE ring.  We leave for future work the question of extending
these methods to cover those cases.

\section{Acknowledgements}

We would like to thank J.~Knapp, L.~Mihalcea,
W.~Xu, H.~Zhang, and H.~Zou for useful discussions.
I.V.M.~was partially supported by the Humboldt Research Award and the Jean d'Alembert Program at the University of Paris--Saclay, as well as the Educational Leave program at James Madison University.  
Part of this work was carried out while I.V.M.~was visiting  the Albert Einstein Institute (Max Planck Institute for Gravitational Physics) as well as LPTHE at Sorbonne Universit\'e, UPMC Paris 06, and he is grateful to both AEI and LPTHE for their generous hospitality.
E.S.~was partially supported by NSF grant PHY-2014086.


\begin{thebibliography}{199}

\addcontentsline{toc}{section}{References}

\bibitem{Witten:1993yc}
E.~Witten,
``Phases of N=2 theories in two-dimensions,''
Nucl. Phys. B \textbf{403} (1993) 159-222,
{\tt arXiv:hep-th/9301042 [hep-th]}.

\bibitem{Morrison:1994fr}
D.~R.~Morrison and M.~R.~Plesser,
``Summing the instantons: Quantum cohomology and mirror symmetry in toric varieties,''
Nucl. Phys. B \textbf{440} (1995) 279-354,
{\tt arXiv:hep-th/9412236 [hep-th]}.

\bibitem{beauville}
A. Beauville, ``Quantum cohomology of complete intersections,''
R.C.P. 25, Vol. 48, Pr\'epubl. Inst. Rech. Math. Av. 1997/42 (1997)
57-68, and Math. Fiz. Anal. Geom. {\bf 2} (1995) 384-398,
{\tt arXiv:alg-geom/9501008 [alg-geom]}.


\bibitem{sheridan} N. Sheridan,
``On the Fukaya category of a Fano hypersurface in projective space,''
Publ. math. de l'IHES {\bf 124} (2016) 165-317,
{\tt arXiv:1306.4143 [math.SG]}.


\bibitem{abpz} H. Arg\"uz, P. Bousseau, R. Pandharipande,
D. Zvonkine, ``Gromov-Witten theory of complete intersections via
nodal invariants,''
J. Topology {\bf 16} (2023) 264-343,
{\tt arXiv:2109.13323 [math.AG]}. 

\bibitem{Collino:1996my}
A.~Collino and M.~Jinzenji,
``On the structure of small quantum cohomology rings for projective hypersurfaces,''
Commun. Math. Phys. \textbf{206} (1999) 157-183,
{\tt arXiv:hep-th/9611053 [hep-th]}.

\bibitem{Melnikov:2005hq}
I.~V.~Melnikov and M.~R.~Plesser,
``The Coulomb branch in gauged linear sigma models,''
JHEP \textbf{06} (2005) 013,
{\tt arXiv:hep-th/0501238 [hep-th]}.

\bibitem{Melnikov:2006kb}
I.~V.~Melnikov and M.~R.~Plesser,
``A-model correlators from the Coulomb branch,''
JHEP \textbf{02} (2006) 044,
{\tt arXiv:hep-th/0507187 [hep-th]}.

\bibitem{arapura} D. Arapura,
{\it Algebraic geometry over the complex numbers},
Springer, 2012.

\bibitem{Bertolini:2013xga}
M.~Bertolini, I.~V.~Melnikov and M.~R.~Plesser,
``Hybrid conformal field theories,''
JHEP \textbf{05} (2014) 043,
{\tt arXiv:1307.7063 [hep-th]}.

\bibitem{Bertolini:2014dia}
M.~Bertolini, I.~V.~Melnikov and M.~R.~Plesser,
``Massless spectrum for hybrid CFTs,''
Proc. Symp. Pure Math. \textbf{88} (2014) 221-230,
{\tt arXiv:1402.1751 [hep-th]}.

\bibitem{Bertolini:2018now}
M.~Bertolini and M.~Romo, ``{Aspects of (2,2) and (0,2) hybrid models},''
{\tt arXiv:1801.04100 [hep-th]}.


\bibitem{Guo:2021aqj}
J.~Guo and M.~Romo, ``{Hybrid models for homological projective duals and
  noncommutative resolutions},''
Lett. Math. Phys.
  {\bfseries 112} (2022)  117,
{\tt arXiv:2111.00025 [hep-th]}.


\bibitem{Erkinger:2022sqs}
D.~Erkinger and J.~Knapp, ``{On genus-0 invariants of Calabi-Yau hybrid
  models},'' 
{\tt arXiv:2210.01226 [hep-th]}.

\bibitem{Vafa:1989xc}
C.~Vafa,
``String vacua and orbifoldized L-G models,''
Mod. Phys. Lett. A \textbf{4} (1989) 1169-1185.

\bibitem{Intriligator:1990ua}
K.~A.~Intriligator and C.~Vafa,
``Landau-Ginzburg orbifolds,''
Nucl. Phys. B \textbf{339} (1990) 95-120.



\bibitem{Kachru:1993pg}
S.~Kachru and E.~Witten,
``Computing the complete massless spectrum of a Landau-Ginzburg orbifold,''
Nucl. Phys. B \textbf{407} (1993) 637-666,
{\tt arXiv:hep-th/9307038 [hep-th]}.

\bibitem{Witten:1990bs}
E.~Witten, ``{Introduction to cohomological field theories},''
{\em Int. J. Mod. Phys.} {\bfseries A6} (1991) 2775--2792.
%%CITATION = IMPAE,A6,2775;%%.

\bibitem{Dijkgraaf:1990qw}
R.~Dijkgraaf, H.~L. Verlinde, and E.~P. Verlinde, ``{Notes on topological
  string theory and 2-D quantum gravity},''. Based on lectures given at Spring
  School on Strings and Quantum Gravity, Trieste, Italy, Apr 24 - May 2, 1990
  and at Cargese Workshop on Random Surfaces, Quantum Gravity and Strings,
  Cargese, France, May 28 - Jun 1, 1990.

\bibitem{Adams:2023imc}
G.~Adams and I.~V. Melnikov, ``{Marginal deformations of Calabi-Yau
  hypersurface hybrids with (2,2) supersymmetry},''
{\tt arXiv:2305.05971 [hep-th]}.


\bibitem{Martinec:1988zu}
E.~J. Martinec, ``{Algebraic geometry and effective lagrangians},''
Phys. Lett. {\bfseries B217} (1989) 431-437.

\bibitem{Vafa:1988uu}
C.~Vafa and N.~P. Warner, ``{Catastrophes and the classification of conformal
theories},''
Phys. Lett. {\bfseries B218} (1989) 51-58. 

\bibitem{Wen:1985qj}
X.~G.~Wen and E.~Witten,
``Electric and magnetic charges in superstring models,''
Nucl. Phys. B \textbf{261} (1985) 651-677.

\bibitem{MR2896292}
V.~I. Arnold, S.~M. Gusein-Zade, and A.~N. Varchenko, {\em Singularities of
  differentiable maps, {V}olume 1},
Modern Birkh\"{a}user Classics, Birkh\"{a}user/Springer, New York,
  2012.



\bibitem{Gu:2020zpg}
W.~Gu, L.~Mihalcea, E.~Sharpe and H.~Zou,
``Quantum K theory of symplectic Grassmannians,''
J. Geom. Phys. \textbf{177} (2022) 104548,
{\tt arXiv:2008.04909 [hep-th]}.

\bibitem{Hellerman:2006zs}
S.~Hellerman, A.~Henriques, T.~Pantev, E.~Sharpe and M.~Ando,
``Cluster decomposition, T-duality, and gerby CFT's,''
Adv. Theor. Math. Phys. \textbf{11} (2007) 751-818,
{\tt arXiv:hep-th/0606034 [hep-th]}.

\bibitem{Caldararu:2007tc}
A.~C\u{a}ld\u{a}raru, J.~Distler, S.~Hellerman, T.~Pantev and E.~Sharpe,
``Non-birational twisted derived equivalences in abelian GLSMs,''
Commun. Math. Phys. \textbf{294} (2010) 605-645,
{\tt arXiv:0709.3855 [hep-th]}.

\bibitem{Hori:2011pd}
K.~Hori,
``Duality in two-dimensional (2,2) supersymmetric non-abelian gauge theories,''
JHEP \textbf{10} (2013) 121,
{\tt arXiv:1104.2853 [hep-th]}.

\bibitem{Bullimore:2014awa}
M.~Bullimore, H.~C.~Kim and P.~Koroteev,
``Defects and quantum Seiberg-Witten geometry,''
JHEP \textbf{05} (2015) 095,
{\tt arXiv:1412.6081 [hep-th]}.



\bibitem{Jockers:2018sfl}
H.~Jockers and P.~Mayr,
``A 3d gauge theory/quantum K-theory correspondence,''
Adv. Theor. Math. Phys. \textbf{24} (2020) 327-457,
{\tt arXiv:1808.02040 [hep-th]}.

\bibitem{Jockers:2019wjh} 
H.~Jockers and P.~Mayr,
``Quantum K-theory of Calabi-Yau manifolds,''
JHEP \textbf{11} (2019) 011,
{\tt arXiv:1905.03548 [hep-th]}.

\bibitem{Jockers:2019lwe}
H.~Jockers, P.~Mayr, U.~Ninad and A.~Tabler,
``Wilson loop algebras and quantum K-theory for Grassmannians,''
JHEP \textbf{10} (2020) 036,
{\tt arXiv:1911.13286 [hep-th]}.


\bibitem{Ueda:2019qhg}
K.~Ueda and Y.~Yoshida,
``3d $ \mathcal{N} $ = 2 Chern-Simons-matter theory, Bethe ansatz, and quantum $K$-theory of Grassmannians,''
JHEP \textbf{08} (2020) 157,
{\tt arXiv:1912.03792 [hep-th]}.


\bibitem{Gu:2022yvj}
W.~Gu, L.~C.~Mihalcea, E.~Sharpe and H.~Zou,
``Quantum K theory of Grassmannians, Wilson line operators, and Schur bundles,''
{\tt arXiv:2208.01091 [math.AG]}.



\bibitem{Gu:2023tcv}
W.~Gu, L.~Mihalcea, E.~Sharpe, W.~Xu, H.~Zhang and H.~Zou,
``Quantum K theory rings of partial flag manifolds,''
{\tt arXiv:2306.11094 [hep-th]}.


\bibitem{Gu:2021yek}
W.~Gu, D.~Pei and M.~Zhang,
``On phases of 3d N=2 Chern-Simons-matter theories,''
Nucl. Phys. B \textbf{973} (2021), 115604
{\tt arXiv:2105.02247 [hep-th]}.



\bibitem{Katz:2004nn}
S.~H.~Katz and E.~Sharpe,
``Notes on certain (0,2) correlation functions,''
Commun. Math. Phys. \textbf{262} (2006) 611-644,
{\tt arXiv:hep-th/0406226 [hep-th]}.

\bibitem{McOrist:2007kp}
J.~McOrist and I.~V.~Melnikov,
``Half-twisted correlators from the Coulomb branch,''
JHEP \textbf{04} (2008) 071,
{\tt arXiv:0712.3272 [hep-th]}.

\bibitem{McOrist:2008ji}
J.~McOrist and I.~V.~Melnikov,
``Summing the instantons in half-twisted linear sigma models,''
JHEP \textbf{02} (2009) 026,
{\tt arXiv:0810.0012 [hep-th]}.

\bibitem{Donagi:2011va}
R.~Donagi, J.~Guffin, S.~Katz and E.~Sharpe,
``Physical aspects of quantum sheaf cohomology for deformations of tangent bundles of toric varieties,''
Adv. Theor. Math. Phys. \textbf{17} (2013)  1255-1301,
{\tt arXiv:1110.3752 [hep-th]}.

\bibitem{Donagi:2011uz}
R.~Donagi, J.~Guffin, S.~Katz and E.~Sharpe,
``A mathematical theory of quantum sheaf cohomology,''
Asian J. Math. \textbf{18} (2014) 387-418,
{\tt arXiv:1110.3751 [math.AG]}.

\bibitem{Guo:2015caf}
J.~Guo, Z.~Lu and E.~Sharpe,
``Quantum sheaf cohomology on Grassmannians,''
Commun. Math. Phys. \textbf{352} (2017)  135-184,
{\tt arXiv:1512.08586 [hep-th]}.

\bibitem{Guo:2016suk} 
J.~Guo, Z.~Lu and E.~Sharpe,
``Classical sheaf cohomology rings on Grassmannians,''
J. Algebra \textbf{486} (2017) 246-287,
{\tt arXiv:1605.01410 [math.AG]}.

\bibitem{Guo:2018iyr}
J.~Guo,
``Quantum sheaf cohomology and duality of flag manifolds,''
Commun. Math. Phys. \textbf{374} (2019) 661-688,
{\tt arXiv:1808.00716 [hep-th]}.

\bibitem{McOrist:2010ae}
J.~McOrist,
``The revival of (0,2) linear sigma models,''
Int. J. Mod. Phys. A \textbf{26} (2011) 1-41,
{\tt arXiv:1010.4667 [hep-th]}.

\bibitem{Guffin:2011mx}
J.~Guffin,
``Quantum sheaf cohomology, a precis,''
Mat. Contemp. \textbf{41} (2012) 17-26,
{\tt arXiv:1101.1305 [math.AG]}.


\bibitem{Melnikov:2012hk}
I.~Melnikov, S.~Sethi and E.~Sharpe,
``Recent developments in (0,2) mirror symmetry,''
SIGMA \textbf{8} (2012) 068,
{\tt arXiv:1209.1134 [hep-th]}.


\bibitem{hartshorne} R. Hartshorne,
{\it Algebraic geometry},
Springer-Verlag, New York, 1977.

\bibitem{Guo:2015gha}
J.~Guo, B.~Jia and E.~Sharpe,
``Chiral operators in two-dimensional (0,2) theories and a test of triality,''
JHEP \textbf{06} (2015) 201,
{\tt arXiv:1501.00987 [hep-th]}.




\bibitem{Closset:2015rna}
C.~Closset, S.~Cremonesi and D.~S.~Park,
``The equivariant A-twist and gauged linear sigma models on the two-sphere,''
JHEP \textbf{06} (2015) 076,
{\tt arXiv:1504.06308 [hep-th]}.







\end{thebibliography}
\end{document}